\documentclass[%
 reprint,
nofootinbib,
 amsmath,amssymb,
 aps,
prd,
showkeys,
]{revtex4-2}

\usepackage{verbatim}

\usepackage{graphicx}
\usepackage{dcolumn}
\usepackage{bm}
\usepackage{hyperref}
\hypersetup{
    colorlinks=true,
    linkcolor=blue,
    filecolor=magenta,      
    urlcolor=cyan
}
\usepackage{color,soul}

\usepackage{soul}
\usepackage{color}
\usepackage{ulem}
\usepackage{slashed}
\usepackage{appendix}

\newcommand{\be}{\begin{equation}}
\newcommand{\ee}{\end{equation}}

\begin{document}


\title{Femtoscopy of $DN$ and $\bar{D}N$ systems}

\author{Mikel F. Barbat}
 \email{mikel.fernandez@ific.uv.es}
 
\affiliation{Instituto de F\'isica Corpuscular (centro mixto CSIC-UV), Institutos de Investigaci\'on de Paterna, C/Catedr\'atico Jos\'e Beltr\'an 2, E-46980 Paterna, Valencia, Spain}

\author{Juan M. Torres-Rincon}
 \email{torres@fqa.ub.edu}
 
\affiliation{Departament de F\'isica Qu\`antica i Astrof\'isica and Institut de Ci\`encies del Cosmos (ICCUB), Facultat de F\'isica, Universitat de Barcelona, Mart\'i i Franqu\`es 1, 08028 Barcelona, Spain}

\author{\`Angels Ramos}
 \email{ramos@fqa.ub.edu}
 
\affiliation{Departament de Fı\'isica Qu\`antica i Astrof\'isica and Institut de Ci\`encies del Cosmos (ICCUB), Facultat de F\'isica, Universitat de Barcelona, Mart\'i i Franqu\`es 1, 08028 Barcelona, Spain}

\author{Laura Tolos}
 \email{tolos@ice.csic.es}
\affiliation{Institute of Space Sciences (ICE, CSIC), Campus UAB, Carrer de Can Magrans, 08193 Barcelona, Spain}
\affiliation{Institut d'Estudis Espacials de Catalunya (IEEC), 08860 Castelldefels (Barcelona), Spain}

\date{\today}

\begin{abstract}
The capability of the ALICE@LHC and STAR@RHIC experiments to reconstruct $D$ mesons has enabled femtoscopic correlation measurements of open-charm mesons in both small and large systems. In this work, we present a theoretical calculation of the correlation functions of $D$ and $\bar{D}$ mesons with nucleons, based on the Koonin-Pratt formalism. We employ an effective Lagrangian to model the interaction between charmed mesons and baryons and apply the TROY formalism to obtain the off-shell $T$-matrix in coupled channels, incorporating the effect of the Coulomb interaction when the pair involves two charged particles. The resulting full coupled-channel wave function is inserted into the Koonin-Pratt equation with channel weights derived from a thermal model.
Additionally, we compute the correlation functions using the Lednický-Lyuboshitz approximation with low-energy scattering parameters extracted from the unitarized amplitudes. We compare these two approaches and provide predictions for different correlated pairs. Our results can be tested against current and future experimental data from the ALICE and STAR collaborations in both proton-proton and heavy-ion collisions.
\end{abstract}

\keywords{Femtoscopy, heavy mesons, effective field theories, $T$-matrix equation, relativistic scattering theory}
\maketitle

\section{Introduction}

Quantum Chromodynamics (QCD) is known to be the fundamental theory of the strong interaction. Many processes at high energies (or shorter distances than the size of the nucleon) can be determined using a perturbative approach involving quark and gluon interactions. However, the low-energy QCD regime is strongly coupled and the interactions among hadrons, the relevant degrees of freedom at this energy scale, are not well determined and usually difficult to analyze experimentally. 

Over the past years, femtoscopy has turned out to be a promising tool to study the interactions among hadrons (see the review of Ref.~\cite{Fabbietti:2020bfg}). Femtoscopy is based on determining hadron-hadron correlations in momentum space as the ratio between the distribution of relative momenta for pairs produced in the same collision and in different collisions.  Since the pioneer works in \cite{HADES:2016dyd,Shapoval:2014yha} and further developments of the STAR Collaboration \cite{STAR:2014dcy,STAR:2015kha,STAR:2018uho}, the
ALICE Collaboration has taken the leading role and determined the correlations for different hadron-hadron systems (see, for example, the works in \cite{ALICE:2017jto,ALICE:2017iga,ALICE:2018nnl,ALICE:2018ysd,ALICE:2019igo,ALICE:2019eol,ALICE:2019gcn,ALICE:2019hdt,ALICE:2019buq,ALICE:2020wvi,ALICE:2020mfd,ALICE:2020mkb,ALICE:2021njx,ALICE:2021ovd,ALICE:2021cyj,ALICE:2021cpv,ALICE:2022yyh,ALICE:2022uso,ALICE:2022mxo,ALICE:2023wjz}).

In the past years the ALICE Collaboration has started investigating hadron-hadron interactions in the charm sector using femtoscopic techniques. More precisely, the correlation functions  of $D^{(\pm)} \pi^{(\pm)}$ and $D^{(\pm)} K^{(\pm)}$ \cite{ALICE:2024bhk} as well as $D^- p$ and $D^+ \bar p$ \cite{ALICE:2022enj} in high-multiplicity pp collisions at 13 TeV have been recently obtained, thus giving new insights into interactions involving $D$ mesons.

Motivated by these recent experimental developments on femtoscopy, several theoretical analyses have been carried out in the charm sector. Some of them have focused on the correlation functions of exotic hidden charmed meson states generated dynamically via the scattering of open-charm mesons, such as $X(3872)$ \cite{Kamiya:2022thy,Abreu:2025jqy}, $T_{cc}$ \cite{Kamiya:2022thy,Vidana:2023olz,Albaladejo:2024lam}, $Z_c$(3900) and $Z_{cs}(3985)$ \cite{Liu:2024nac}, as well as the correlation functions of $P_c(4440)$ and $P_c(4457)$ that emerge from the interaction of $D$ mesons with baryons \cite{Liu:2023wfo}. Also, some recent theoretical studies have aimed at determining the correlation functions of open-charm mesons with light mesons \cite{Albaladejo:2023pzq,Torres-Rincon:2023qll,Liu:2023uly,Ikeno:2023ojl,Khemchandani:2023xup}, and the charmonium-nucleon \cite{Liu:2025oar} or even the $\alpha$-charmonium \cite{Etminan:2025tiy} correlation functions.

In view of the recent experimental results on femtoscopy for $D$ mesons with protons \cite{ALICE:2022enj} and following our previous analysis of the $D \pi$ correlations functions \cite{Torres-Rincon:2023qll}, in the present work we focus on the determination of the correlation functions for $D^+p$, $D^-p$, $D^0p$ and $\bar D^0 p$ systems. For that purpose we start from the interaction of $D$ and $\bar D$ mesons with nucleons obtained from the vector meson-exchange model of Ref.~\cite{Hofmann:2005sw} in the zero-range approximation~\cite{Jimenez-Tejero:2009cyn}. The correlation functions are then calculated using the off-shell $T$-matrix TROY approach recently developed in~\cite{Torres-Rincon:2023qll}. We also compare our results with those obtained within the Lednick\'y-Lyuboshitz (LL) approximation in order to assess the differences and caveats of this approximation. Within the simple LL approximation, we employ different $D-$ and $\bar D-$nucleon interactions \cite{Hofmann:2005sw,Lutz:2005vx, Mizutani:2006vq,Garcia-Recio:2008rjt,Haidenbauer:2010ch, Haidenbauer:2007jq,
Fontoura:2012mz,Yamaguchi:2011xb}, so as to determine the dependence of the $D^+p$, $D^-p$, $D^0p$ and $\bar D^0 p$ correlation functions on the underlying interaction model.

In Sec.~\ref{sec:formalism} we provide theoretical details of the interaction model between charmed mesons and nucleons, the unitarization techniques to obtain the $T$-matrices and the wave functions, and the femtoscopy correlation functions. In Sec.~\ref{sec:results} we present the results of the correlation functions using the LL approximation as well as the off-shell approach using the TROY formalism. In Sec.~\ref{sec:Conclusions} we give a summary of this work.

\section{Formalism~\label{sec:formalism}}

In this section, we provide theoretical details to compute the femtoscopy correlation functions for $DN$ and $\bar{D}N$ pairs. In Sec.~\ref{sec:strong} we present the model that describes the strong interaction between the hadrons. In Sec.~\ref{sec:Tmatrix} we provide details on how the $T$-matrix equation is solved both in the off-shell case and in the on-shell factorization approach. In Sec.~\ref{sec:correlation}, we introduce the Koonin-Pratt (KP) equation as well as the two approaches we consider: the LL approximation based on the effective-range expansion of the on-shell amplitudes and the off-shell calculation within TROY.

\subsection{Meson-baryon strong interaction~\label{sec:strong}}

An effective description of meson-baryon scattering involving charm and anticharm degrees of freedom is developed in Ref.~\cite{Hofmann:2005sw}. Using that approach, we focus on the $C=1$ and $C=-1$ sectors, to accommodate the $DN$ and $\bar{D}N$ channels, respectively. We follow the notation and conventions given in Ref.~\cite{Jimenez-Tejero:2009cyn}.

The lowest-order tree-level diagrams of the meson–baryon interaction are composed by vector meson exchanges~\cite{Jimenez-Tejero:2009cyn}. The $s$-wave amplitude $V_{ij}$ (where $i$ and $j$ refer to the incoming and outgoing meson-baryon channels) is  dominated by the $l=0$ projection of the $t$-channel meson exchange, while the $u$- and $s$-channel contributions only become more relevant in the $p$-wave amplitude, potentially playing a significant role at higher energies. In fact, in Ref.~\cite{Oller:2000fj} these contributions were evaluated for the $\bar{K}N$ system and were found to amount to approximately 20\% of the $t$-channel contribution at energies around 200 MeV above threshold. In the charm sector, such contributions are expected to be even smaller, as discussed in Ref.~\cite{Jimenez-Tejero:2009cyn}.

This $t$-channel vector meson exchange contribution can be computed using the Lagrangian defined in Ref.~\cite{Lutz:2005vx,Hofmann:2005sw},  including all allowed vector-meson exchanges in a given  transition $i\to j$. From Ref.~\cite{Jimenez-Tejero:2009cyn}, the $t$-channel exchanged potential reads,
\begin{widetext}
\be 
V^{ISC}_{ij} (q_i,p_i,q_j,p_j) =\frac{g^2}{4} \bar{u}(p_j) \sum_v \frac{C^{ISC}_{ij,v}}{t-m_v^2} \left[  \slashed{q}_i + \slashed{q}_j - \frac{q_i^2-q_j^2}{m_v^2} (\slashed{q}_i - \slashed{q}_j) \right]  u(p_i) \ ,
\ee
\end{widetext}
where $g$ is the universal vector meson coupling constant, $C^{ISC}_{ij,v}$ represents the isospin coefficients depending on the ($ISC$) quantum numbers as well as the $i$ and $j$ channels and the vector meson exchanged $v$~(see~\cite{Hofmann:2005sw} for explicit values). The 4-momenta $q_i$ and $q_j$ correspond to the incident and outgoing mesons. Finally, $u(p_i),\bar{u}(p_j),p_i,p_j$ are the incident and outgoing baryon wave functions, as well as their respective momenta.

Then, we assume the $|t| \ll m_v^2$ limit to simplify the $t$-dependence in the denominator as well as the third term in the brackets. This is equivalent to making a contact interaction at the vertex. As explained in Ref.~\cite{Jimenez-Tejero:2009cyn}, the error introduced in this approximation is small. Finally, we take the $s$-wave projection of this so-called Zero-Range (ZR) approximation for the interaction kernel,
\begin{widetext}
\be
V_{ij,  \ \textrm{ZR}}^{ISC} (|\bm{q}_i|=|\bm{p}_i|,|\bm{q}_j|=|\bm{p}_j|; \sqrt{s})  =-\dfrac{N}{4} g^2 \sum_v \dfrac{C^{ISC}_{ij,v}}{m_v^2}  \left( 2 \sqrt{s}-E_i-E_j + \frac{p_i^2}{E_i+M_i} + \frac{p_j^2}{E_j+M_j} \right) \ ,
    \label{eq:VZR}
\ee
\end{widetext}
where, in the center of mass (CM) frame,
\be
N = \sqrt{\dfrac{M_i+E_i}{2M_i}} \ \sqrt{\dfrac{M_j+E_j}{2M_j}} \ , 
\ee
is a normalization factor arising from the Dirac spinors;  and $M_i$, $M_j$ and $E_i=(s+M_i^2-m_i^2)/(2\sqrt{s})$, $E_j=(s+M_j^2-m_j^2)/(2\sqrt{s})$ are the masses and energies of the baryons in channels $i$ and $j$, respectively. Equation~\eqref{eq:VZR} serves as an off-shell kernel for the meson-baryon scattering in the ZR approximation. The on-shell version of~\cite{Jimenez-Tejero:2009cyn} is simply achieved by setting $p_{i/j}^2=E_{i/j}^2-M_{i/j}^2$. In this work, both off-shell and on-shell versions of this potential in the ZR approximation will be used.

Finally, by assuming a common vector-meson mass, one can apply the Kawarabayashi-Suzuki-Fayyazuddin-Riazuddin rule~\cite{Kawarabayashi:1966kd,Riazuddin:1966sw}, $g = m_v/(\sqrt{2} f)$, where $f$ is the meson weak decay constant. In this case,
the potential takes the form of the Weinberg–Tomozawa (WT) interaction, 
\begin{equation}
    V_{ij, \ \textrm{WT}}^{ISC} (\sqrt{s})= -\dfrac{N}{8f^2} C_{ij}^{\textrm{WT}} \ (2\sqrt{s} - M_i -M_j) \ ,
    \label{eq:VWT}
\end{equation}
where the new isospin coefficients,
\begin{equation}
    C_{ij}^{\textrm{WT}}=\kappa \sum_v C^{ISC}_{ij,v} \ ,
\end{equation}
carry a correction factor $\kappa = 1/4$ only when the exchanged vector meson contains a charm quark, to balance from the universal vector mass assumption \cite{Jimenez-Tejero:2009cyn}.

In this work, we focus on four isospin/strangess/charm sectors. To account for $D^0p$ and $D^+p$ cases we will consider the $\{ISC\}=\{0,0,1\}$ $DN$ sector with 7 coupled channels,
\begin{equation}\pi \Sigma_c, DN, \eta \Lambda_c, K\Xi_c, K\Xi'_c, D_s \Lambda, \eta' \Lambda_c \ , 
\end{equation}
leaving aside heavier channels with additional $c\bar{c}$ content, and also the $\{ISC\}=\{1,0,1\}$ $DN$ sector with 8 coupled channels,
\begin{equation}
 \pi \Lambda_c, \pi \Sigma_c, DN, K\Xi_c, \eta \Sigma_c, K\Xi'_c, D_s \Sigma, \eta' \Sigma_c\ . 
\end{equation}

For the $D^-p$ and $\bar{D}^0 p$ cases we work with the $\{ISC\}=\{0,0,-1\}$ and $\{ISC\}=\{1,0,-1\}$ $\bar{D}N$ sectors, each of them having a single channel.

We anticipate that to compute the correlation functions with TROY, it is more convenient to work in the physical basis (with the physical mass of each state). In this case, the number of coupled channels is 16, 9, 2 and 1, for the cases $D^0p, D^+p, D^-p, \bar{D}^0p$, respectively. \\

\subsection{Off- and on-shell $T$-matrices}
\label{sec:Tmatrix}

The perturbative potential $V_{ij}(q',q; \sqrt{s})$ is used as the kernel of the Bethe-Salpeter (or $T$-matrix) equation,
\begin{widetext}
\be
T_{if} (q',q; \sqrt{s})  = V_{if} (q',q; \sqrt{s})   + \sum_l \int \frac{d^3k}{(2\pi)^3}
\frac{2M_l (E_l+\omega_l)}{2 E_{l} \omega_l} \frac{V_{il} (q',k;\sqrt{s}) T_{lf} (k,q; \sqrt{s})}{s-(E_{l}+\omega_{l})^2 + \mathrm{i} \eta} \ , \label{eq:Tmatrix} 
\ee
\end{widetext}
where we denote the initial and final CM momenta as $q'=|\bm{q}_i|$ and $q=|\bm{q}_f|$, respectively.  The quantities $E_{l}=\displaystyle\sqrt{M_l^2+k^2}$ and $\omega_l=\displaystyle\sqrt{m_l^2+k^2}$ are the intermediate baryon and meson energies, respectively. The on-shell CM momentum of the final ($f$) baryon-meson pair is $q=\lambda^{1/2}(s,M_f^2,m_f^2)/(2\sqrt{s})$, where $\lambda$ is the K\"all\'en function. The initial CM momentum of the pair $q'$ is not fixed by $\sqrt{s}$ since in Eq.~\eqref{eq:Tmatrix} we consider the half off-shell $T$-matrix elements.

In the on-shell approximation, $q'$ is taken to be $q'=\lambda^{1/2}(s,M_i^2,m_i^2)/(2\sqrt{s})$ and both $V_{il},T_{lf}$ can be pulled out of the integral. One gets
\be T_{if} (\sqrt{s}) = V_{if}(\sqrt{s}) + \sum_l   V_{il} (\sqrt{s}) G_l(\sqrt{s}) T_{lf} (\sqrt{s}) \ , \label{eq:os} \ee
with
\be G_l(\sqrt{s}) =  2M_l \int \frac{d^3k}{(2\pi)^3} \frac{E_l+\omega_l}{2 E_l \omega_l} \frac{1}{s-(E_l+\omega_l)^2 + \mathrm{i} \eta} \ \label{eq:gloop} . \ee
Both this loop integral and that of Eq.~(\ref{eq:Tmatrix}) exhibit an ultraviolet (UV) divergence, which we will regularize by introducing a UV 3-momentum cutoff, $k_{\textrm{max}}$, whose value will be determined by imposing the position of the dynamically generated $\Lambda_c(1295)$ in the $DN$ sector. The same value will be used in the $\bar{D}N$ sector

\subsection{Femtoscopy correlation function~\label{sec:correlation}}

The primary quantity observed in femtoscopy is the correlation function of a given channel $f$. It is expressed as a function of the reduced relative momentum of two particles in their rest frame, determined using the KP formula~\cite{KOONIN197743,PhysRevC.42.2646},
\begin{equation}
    C(k)=\int d^3 r \ S(r) |\Psi_f(\bm{k}; \bm{r})|^2 \ , \label{eq:KP}
\end{equation}
where $S(r)$ is the source function, describing the spatial emission of particles, for which we assume a Gaussian profile,
\begin{equation}
    S(r)= (4\pi r_0^2)^{-3/2} \ \exp\left( -\dfrac{r^2}{4 r_0^2} \right) \ ,
\end{equation}
where $r_0$ defines the size of the source. In Eq.~\eqref{eq:KP}, $\Psi_f(\bm{k}; \bm{r})$ is the wave function of the two-particle system, which encodes information about their interaction. It can be addressed by solving the appropriate Schr\"odinger equation or, as we do in this work, using the Lippmann-Schwinger equation after solving the $T$-matrix equation~\eqref{eq:Tmatrix}.

\subsubsection{Lednick\'y-Lyuboshitz approximation}

A simple approximation for computing the correlation function is given in Ref.~\cite{Gmitro:1986ay}, where the wave function is taken to be the asymptotic one, even for $r\rightarrow 0$, with a combination of the incident plane wave plus the spherical outgoing wave modulated by the scattering amplitude $f(k)$. The latter is usually taken in the effective-range expansion, whose inverse follows an expansion in powers of the relative momentum $k$. When the Coulomb effects are included, the correlation function in the LL model reads~\cite{Torres-Rincon:2024znb,Abreu:2025jqy},
\begin{widetext}
\be 
   C^\textrm{C}_{\textrm{LL}}(k)  = A_{\textrm{C}}(k) \left\{ 1+ \dfrac{|f_\textrm{C}(q)|^2}{4 r_0^2} \left[ 1+e^{-(2kr_0)^2} 
     + A_\textrm{C}^2 (1-e^{-(2kr_0)^2})-\dfrac{d_0}{\sqrt{\pi}r_0}\right] + 
    \dfrac{2 \ {\rm Re} f_\textrm{C}(k)}{\sqrt{\pi} r_0} F_1(2kr_0)-\dfrac{A_\textrm{C} \ {\rm Im}f_\textrm{C}(k)}{r_0} F_2(2kr_0) \right\} \ , 
    \label{eq:CLLC}
\ee
\end{widetext}
where $F_1(x)$ and $F_2(x)$ are functions defined as,
\begin{equation}
    F_1(x)= \dfrac{1}{x} \int_0^x dt \ e^{(t^2-x^2)}, \qquad F_2(x)=\dfrac{1-e^{-x^2}}{x}.
\end{equation}
The functions $A_\textrm{C}(q)$ and $f_\textrm{C}(q)$ are the Gamow factor and the scattering amplitude with Coulomb effects, respectively,
\begin{align}
    A_\textrm{C}(k)&= \dfrac{2 \pi \gamma}{ {\rm{exp}} \left( 2 \pi \gamma \right) -1}  \ , \\[10pt]
    f_\textrm{C}(k)&=\left( \dfrac{1}{a_0}+\dfrac{1}{2}d_0 k^2-2k \gamma h( \gamma^{-1})-i k A_{\textrm{C}}(k)  \right)^{-1} \ ,
    \label{eq:Gamow}
\end{align}
where $\gamma=Z_1 Z_2 \mu \alpha/k$ is the Sommerfeld factor. The parameters $a_0$ and $d_0$, stand for the scattering length and effective range, respectively, and determine the low-energy expansion of the strong-only scattering amplitude. In our convention, $a_0>0$ indicates attraction, while $a_0<0$ indicates repulsion or the presence of a bound state. Finally, the function $h(x)$ is defined as,
\begin{equation}
    h(\gamma^{-1})= -\log (|\gamma|)+ \frac12 \psi(1-i\gamma)+ \frac12 \psi(1+i\gamma) \ , 
\end{equation}
where $\psi(z)$ is the digamma function. In the absence of Coulomb interaction, on sets $\gamma \rightarrow 0$ and recovers the standard LL correlation function~\cite{lednicky1982effect},

\begin{align}
    C_{\textrm{LL}}(k)&=1+\dfrac{1}{2} \left| \dfrac{f(k)}{r_0} \right| ^2 \left(1-\dfrac{d_0}{2 \sqrt{\pi} r_0} \right)  \nonumber \\[10pt]
    &+ \dfrac{2 {\rm Re} f(k)}{\sqrt{\pi} r_0} F_1(2kr_0) -\dfrac{ {\rm Im}f(k)}{r_0} F_2(2kr_0) \ ,
    \label{eq:CLL}
\end{align}
where $f(k)$ is the $\gamma=0$ limit of $f_{\textrm{C}}(k)$.

It is important to note that the LL approximation does not account for the effects of coupled channels. Although the computation of the $T$ matrix does include them, the scattering parameters used in the formulas are obtained only for the direct (elastic) channel.

\subsubsection{Off-shell $T$-matrix (TROY) approach~\label{sec:TROY}}

The TROY (T-matrix-based Routine for hadrOn femtoscopY) formalism uses the solution of the half off-shell $T$-matrix equation (\ref{eq:Tmatrix}) to construct the scattering pair wave function~\cite{Torres-Rincon:2023qll}.

Since the $T$-matrix equation contains all channels coupled to the one that is observed, one needs a more general KP formula,
\be 
C(k) = \int d^3r \sum_{i} w_{i} \ S(r) |\Psi_{if} (\bm{k};\bm{r})|^2 \ , \label{eq:KP2}
\ee
where the weights $w_i$ measure the production probability of the different pair combinations at freeze-out. 

The calculation of $w_i$ requires a model for thermal production of hadronic states, plus feed-down corrections for those states that populate the final hadrons beyond the freeze-out. Therefore, it can be based on a statistical thermal model, such as Thermal-FIST~\cite{Vovchenko:2019pjl}, and needs information of the type of collision at hand (centrality, collision energy, kinematic cuts of the detected particles, efficiency of measured yields, etc.). It also needs to know the kinematic distribution of the particles that can lead to a pair correlation with a given relative momentum, see e.g.~\cite{Encarnacion:2024jge}.

Leaving for a future study a more complete model for $w_i$, we adopt here a static fireball model and compute the thermal multiplicity of each species as~\cite{Letessier:2022fax}
\be \frac{dn}{m_\perp dm_\perp dy d\varphi} \sim m_\perp \cosh{y} \exp \left( - m_\perp \cosh y/T \right) \ ,
\label{eq:thermal_mult}
\ee
where $m_\perp=\sqrt{m^2+p_\perp^2}$, $m$ is the mass of the particle, $p_\perp$ its transverse momentum, $y$ is the rapidity, $\varphi$ is the azimuthal angle of the particle's momentum, and $T$ is the temperature. We integrate this multiplicity without kinematic restrictions and use $T=171$ MeV for high-multiplicity pp collisions at $\sqrt{s}=13$ TeV at the LHC~\cite{ALICE:2022yyh}. Then, the thermal weight $w_i$ for a given pair---using $w_f=1$ as the reference value---is given by
\be w_i = \frac{n_{1,i} \times n_{2,i}}{ n_{1,f} \times n_{2,f} } \ , 
\label{eq:weight}
\ee
where $1,2$ denote the first and second particles of the pair. This prescription, while simple, is able to provide the correct order of magnitude of the weights. In particular, the pairs affected by feed down can deviate more from the actual weight; however, this is a better approach than setting to unity all weights.  We have checked that the weights $w_i$ obtained by our prescription do not deviate much from those of the recent work~\cite{Encarnacion:2024jge} for the particular systems explored there.

Turning back to Eq.~\eqref{eq:KP2}, the total wave function is splitted as,
\be 
 \Psi_{if} (\bm{k};\bm{r})= \delta_{if} [\Psi^{\textrm{C}}_{if} (\bm{k};\bm{r}) -\Psi^{\textrm{C}}_{0,if}(kr)] + \varphi_{if}(k;r) \ , \label{eq:WF}
\ee
where $\Psi^{\textrm{C}}_{if} (\bm{k};\bm{r})$ is the total Coulomb wave function (or plane wave in the absence of electromagnetic forces),
$\Psi^{\textrm{C}}_{0,if}(kr)$ is its $s$-wave projection and $\varphi_{if} (k;r)$ is the $s$-wave projection of the total wave function (including strong and Coulomb interactions).  In writing Eq.~\eqref{eq:WF} we have assumed that at low energies only the $l=0$   wave is modified by the strong interaction. Specific details on the Coulomb wave functions are given in Ref.~\cite{Torres-Rincon:2023qll}.

After solving Eq.~\eqref{eq:Tmatrix}, TROY computes the interacting wave function out of the off-shell $T$-matrix elements as,
\begin{eqnarray}
\varphi_{if}(k;r) &=&j_0(kr) \delta_{if}  \label{eq:splitted}  \\
& +&  \int \frac{d^3q'}{(2\pi)^3} \frac{2M_i (E_i+\omega_i)}{2E_i \omega_i} \frac{T_{if}(q',k;\sqrt{s}) j_0 (q' r)}{s-(E_i+\omega_i)^2+i\eta} \ , \nonumber  
\end{eqnarray}
where $j_0(x)$ is the spherical Bessel function of zeroth order, and $\sqrt{s}=\sqrt{M_f^2+k^2}+\sqrt{m_f^2+k^2}$. We note that, in the case of two charged particles, the $T$ matrix entering the above equation is solved including 
both the strong and Coulomb interaction kernels, as shown in Ref.~\cite{Torres-Rincon:2023qll} and described in Appendix~\ref{ap:coulomb}.

For comparison, we also compute the correlation function given in Eq.~(\ref{eq:KP2}) using the on-shell approach, where the interacting wave function of Eq.~(\ref{eq:splitted}) is obtained from the on-shell form of the $T$ matrix.

\section{Results}
\label{sec:results}

In this section, we present the correlation function of $DN$ and $\bar{D}N$ pairs, calculated using different approximations. In Sec.~\ref{sec:LL} we employ the LL approach for small system sizes and compare our results with those obtained for other interaction models in the literature. In Sec.~\ref{sec:TROYresults} we apply the off-shell TROY formalism for the same small systems.  In Sec.~\ref{sec:size} we use TROY to compare to ALICE results in pp collisions and predict the correlation functions for larger systems, as those studied by the STAR collaboration.

\subsection{LL calculations~\label{sec:LL}}

In this section we present results for the correlation functions of charged particle pairs, $D^+ p$ and $D^- p$, within the LL approach of Eq.~(\ref{eq:CLLC}). We derive the required low-energy parameters from our on-shell version of the $T$-matrix given in Eq.~\eqref{eq:os}. For the ZR interaction kernel we employ  $g=6.0$ and $k_{\textrm{max}}=674$ MeV, while we use $f=1.15f_\pi$ with $f_\pi=92$ MeV and $k_{\textrm{max}}=752$ MeV in the case of the WT interaction.

First, we present the $DN$ and $\bar{D}N$ scattering lengths in the isospin basis in Tables \ref{tab:DN_iso} and \ref{tab:DbarN_iso}, respectively, which also display the outcome from other works for comparison. 

With regard to the $DN$ system, the $s$-wave $I=0$ and $I=1$ scattering lengths predicted by the WT model and those in the ZR approximation, both in the on-shell case, are very similar. This is an expected result, as the difference between the two approaches arises because of the slightly different energy-dependent $DN$ interaction kernel. In both WT and ZR cases, the $I=0$ 
$\Lambda_c(2595)$ is generated dynamically at its mass by adjusting the cutoff, whereas in the $I=1$ sector a $\Sigma_c$ with a mass close to the $DN$ threshold of 2806 MeV is obtained, thus leading to a scattering length having a large negative real part. Note that the scattering lengths reported in Ref.~\cite{Haidenbauer:2010ch} for the WT model of Ref.~\cite{Mizutani:2006vq} slightly differ from those of the WT model shown here due to the slightly different cutoff used to regularize the loop function. Moreover, we show the scattering lengths of the $DN$ interaction of Ref.~\cite{Hofmann:2005sw}, reported in Ref.~\cite{Lutz:2005vx}.  
In that work, there is no $I=1$ quasi-bound state close to the $DN$ threshold but it lies 180 MeV below, explaining the smaller size of the corresponding scattering length compared to the WT and ZR results shown here. As for the scattering lengths coming from the meson-exchange model of Ref.~\cite{Haidenbauer:2010ch}, they are qualitatively similar in both isospin sectors to those shown for WT and ZR models. This is due to the fact that the meson-exchange model also generates dynamically the $I=0$ $\Lambda_c(2595)$ and the $I=1$ $\Sigma_c(2800)$ for masses close to those obtained in our present study. 
We also show the scattering lengths for the SU(8) model of Ref.~\cite{Garcia-Recio:2008rjt}, reported in Ref.~\cite{Haidenbauer:2010ch}. In that approach, both $I=0$ and $I=1$ scattering lengths are positive, being the $I=0$ one compatible with zero.  The discrepancy with previous works has its origin in the presence of other resonant states beyond the $I=0$ $\Lambda_c(2595)$ and the $I=1$ $\Sigma_c(2800)$ close to the $DN$ threshold, as described in Ref.~\cite{Garcia-Recio:2008rjt}. Finally, we include the results of~\cite{SAKAI2020135623} where the scattering lengths were extracted from experimental data of the $pD^0$ invariant-mass distribution in the three-body decay $\Lambda_b \rightarrow \pi^- p D^0$. While the resulting $I=0$ scattering length covers roughly the range of values obtained by the various theoretical models, the central value of the $I=1$ scattering length turns out to have more sizable real and imaginary parts. More notably, even taking the uncertainties of the analysis of Ref.~\cite{SAKAI2020135623} into account, the real part of the $I=1$ scattering length cannot be reproduced by the models.

In the case of $\bar DN$, the $I=0$  scattering length in the WT approach is zero, as the corresponding interaction is strictly vanishing in this channel. A result compatible with zero is consistently obtained in the ZR approach, as seen in Table~\ref{tab:DbarN_iso}. For $I=1$, both the WT and ZR prescriptions produce a similarly sizable negative scattering length, thus indicating a repulsive $\bar DN$ interaction in that sector. Upon comparing these results with those from Ref.~\cite{Lutz:2005vx}, we observe that,  while the $I=1$ value is very close to the ones for WT and ZR, there is a disagreement in the $I=0$ scattering length, that we are not able to trace back to the regularization method. In Ref.~\cite{Haidenbauer:2007jq} a value close to zero was obtained for the $I=0$ scattering length, whereas a more repulsive scattering length than our results for WT and ZR was determined for $I=1$. This model is based on the exchange of mesons between $\bar D$ and $N$ supplemented by a short-range one-gluon exchange and, in fact, about half of the repulsive scattering length in $I=1$ comes from the hadronic meson-exchange contributions, which can be mapped to a certain extent to the WT interaction used in the present work. This result was revisited in Ref.~\cite{ALICE:2022enj}, by modifying the coupling to the scalar meson, leading to a much more attractive interaction in $I=0$ and an almost negligible one in $I=1$. Furthermore, the analysis of Ref.~\cite{Yamaguchi:2011xb}, with a model that considers a long range one-pion exchange interaction, obtains a bound $\bar DN$ state in $I=0$, leading to a very large and negative $I=0$ scattering length, whereas the $I=1$ one is much smaller
. Notice that the value of Ref.~\cite{Yamaguchi:2011xb} for the $I=0$ scattering length is quoted in~\cite{ALICE:2022enj} with an opposite sign, because the convention in the former paper is contrary to the ALICE one (and ours). The work of Ref.~\cite{Fontoura:2012mz} uses a quark model that confines color and realizes dynamical chiral symmetry breaking, giving rise to values for $I=0$ and $I=1$ closer to the WT and ZR results, as reported in Ref.~\cite{ALICE:2022enj}.

%

\begin{table}[h!]
\centering
\begin{tabular}{lcc}
\hline \hline
$DN$ & $a_0~({I=0})[{\rm fm}]$ & $a_0 ~(I=1)[{\rm fm}]$ \\ \hline
WT & $-0.58$ & $-1.55+\rm{i} 0.75 $ \\
ZR & $-0.63$& $-1.79+ \rm{i} 0.32$\\
Mizutani  
\cite{Mizutani:2006vq,Haidenbauer:2010ch} & $-0.57$ & $-1.47+ \rm{i}0.65$ \\
Hofmann-Lutz \cite{Hofmann:2005sw,Lutz:2005vx} & $-0.43$ & $-0.41$\\
Haidenbauer \cite{Haidenbauer:2010ch} & $-0.41+\rm{i}0.04$ & $-2.07+\rm{i}0.57$ \\
Garcia-Recio 
\cite{Garcia-Recio:2008rjt,Haidenbauer:2010ch} & $(4+\rm{i}2)\cdot 10^{-3}$ & $0.33+\rm{i}0.05$ \\ 
Sakai \cite{SAKAI2020135623} & $-0.79_{-0.61}^{+0.66}$ & $-3.8_{-2.0}^{+1.4}+\rm{i}2.7_{-2.7}^{+1.6}$ \\ 
\hline \hline
\end{tabular}
\caption{Scattering lengths for $DN$ with the Weinberg-Tomozawa (WT) and Zero-Range (ZR) approximations compared with other theoretical models in the literature.}
\label{tab:DN_iso}
\end{table}

\begin{table}[h!]
\centering
\begin{tabular}{lcc}
\hline
\hline
$\bar{D}N$ & $a_0~({I=0})[{\rm fm}]$ & $a_0 ~(I=1)[{\rm fm}]$ \\ \hline
WT & $0$ & $-0.29$ \\
ZR & $0.02$ & $-0.34$ \\
Hofmann-Lutz \cite{Lutz:2005vx} & $-0.16$ & $-0.26$ \\
Haidenbauer \cite{Haidenbauer:2007jq} & $0.07$& $-0.45$\\
Haidenbauer, \\ $g_{\sigma}^2/4 \pi=2.25$  \cite{ALICE:2022enj} & $0.67$ & $0.04$ \\
Yamaguchi \cite{Yamaguchi:2011xb,ALICE:2022enj} & $-4.38$ & $-0.07$ \\
Fontoura \cite{Fontoura:2012mz,ALICE:2022enj} & $0.16$ & $-0.25$ \\ \hline \hline
\end{tabular}
\caption{Scattering length for $\bar{D}N$ with the Weinberg-Tomozawa and Zero-Range approximation compared with other theoretical models.}
\label{tab:DbarN_iso}
\end{table}

We now present the correlation functions for charged particle pairs, using the LL approximation with Coulomb interaction of Eq.~(\ref{eq:CLLC}) and employing a source radius of $r_0=1~ \rm{fm}$. Since the aim is to present a comparison of our results with those obtained for the interaction models of previous studies, which only provide information on the scattering length, our results are obtained setting the scattering range to zero. We have checked that, in the $\bar{D}N$ case, the effect of the scattering range in the LL formula is negligible. 

\begin{figure}[h!]
    \centering
    \includegraphics[scale=0.45]{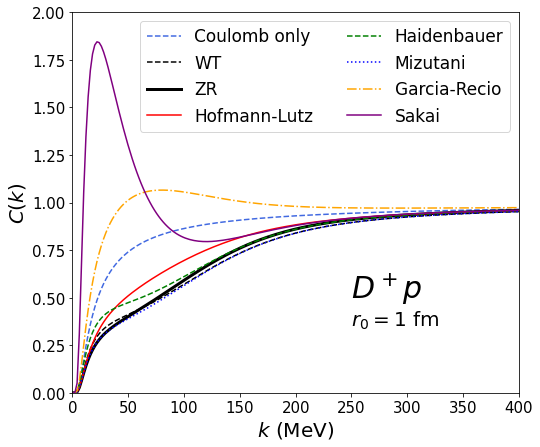}

    \caption{$D^+p$ correlation function as function of the relative momentum. Our results with the ZR approximation are shown in black, in comparison with other works. All of them were calculated using the LL model, Eq.~(\ref{eq:CLLC}), with the low-energy scattering parameters from each paper.}
    \label{fig:CLLD+Comparacion}

\end{figure}

The correlation functions for the purely $I=1$ $D^+p$ system are shown in Fig.~\ref{fig:CLLD+Comparacion}. The repulsive effect caused by the Coulomb interaction, leading to a correlation function below one, dominates the correlation function,  although some differences can be appreciated between the models. The negative sign of the scattering length simulates a repulsive interaction in the majority of models, reducing further the correlation function with respect the the Coulomb only contribution. As mentioned before, the model of Ref.~\cite{Garcia-Recio:2008rjt} presents a positive scattering length, hence an attractive interaction, leading to a higher correlation function compared to the other models and reaching values larger than one for a certain range of momenta. The scattering lengths in~\cite{SAKAI2020135623}, in particular the large value of the $I=1$ channel, produce a correlation function peaked around $k =30$ MeV and above one, signalling the strong coupling of two near-threshold bound states found in that reference, identified to a possible two-pole structure of the $\Sigma_c(2800)$ state.

\begin{figure}[h!]
    \centering
    \includegraphics[scale=0.45]{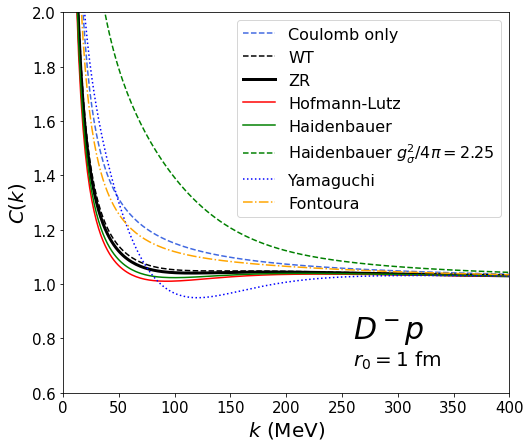}

    \caption{$D^-p$ correlation function as function of the relative momentum. Our results with the ZR approximation are shown in black, in comparison with other works. All of them were calculated using the LL model, Eq.~(\ref{eq:CLLC}), with the low-energy scattering parameters from each paper.}
    \label{fig:CLLD-Comparacion}

\end{figure}

The $D^-p$ correlation functions for different models are shown in Fig.~\ref{fig:CLLD-Comparacion}. In this case, the attractive Coulomb effect dominates, leading in general to values of the correlation function larger than one. Since this system consists of a combination of isospins, the scattering length used is taken as $a=(a_{I=0}+a_{I=1})/2$ for each model. The dispersion of  scattering length values presented earlier results in significantly different correlation functions. The small and negative averaged scattering lengths obtained in this work with the WT and ZR models lead to correlation functions that are very similar to those from other studies. The repulsive interaction associated with these values causes a moderate deviation compared to the dominant attractive Coulomb effect. In contrast, the large negative scattering length reported in Ref.~\cite{Yamaguchi:2011xb} for $I=0$ induces a substantial apparently repulsive effect, as seen in the dotted blue line in Fig.~\ref{fig:CLLD-Comparacion}, where the correlation function is significantly lower than in the other works. This behavior is in fact signalling the presence of a bound state generated by the model of Ref.~\cite{Yamaguchi:2011xb}. The opposite effect is observed in the results revisited in Ref.~\cite{ALICE:2022enj} using the model from Ref.~\cite{Haidenbauer:2007jq}. In this case, the positive and large scattering length indicates an attractive interaction, leading to larger correlation function that agrees with experimental data~\cite{ALICE:2022enj}.

\subsection{TROY framework~\label{sec:TROYresults}}

In this subsection, we present the results obtained using the TROY formalism including all coupled channels in the off-shell $T$-matrix approach. We employ the ZR kernel interaction with $g=6.0$ but adjust the cut-off value to $k_{\textrm{max}}= 630$ MeV so that the $I=0$ amplitude of the $DN$ channel has a maximum at the position of the $\Lambda_c(2595)$ state. 
Then, we apply the same regulator for the remaining $DN$ and ${\bar D}N$ channels. In particular, in the $I=1$ sector we find a maximum of the $DN$ scattering amplitude at $\sqrt{s}=2798$ MeV, which we identify with the $\Sigma_c(2800)$ \cite{ParticleDataGroup:2024cfk}, consistent with the findings of Ref.~\cite{Jimenez-Tejero:2011dif}.

In the following plots, we compare the TROY results (solid black lines) with those of the LL approximation (short-dashed red lines), obtained with the corresponding TROY low-energy parameters (first column in Table~\ref{tab:scat_lengths}). We also show results for the correlation function obtained with the on-shell version of the $T$-matrix (dash-dotted blue lines). Again, we use a small Gaussian source radius, $r_0=1$ fm, to enhance the effect of the strong interaction in the correlation function. 

\begin{figure}[h!]
\centering
\includegraphics[scale=0.55]{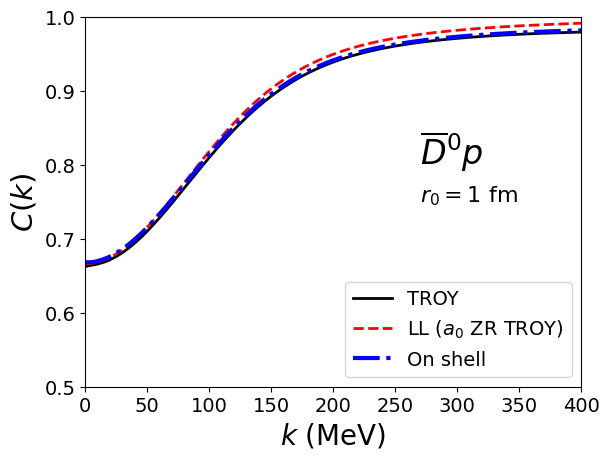}
\caption{$\bar{D}^0 p$ correlation function as function of the relative momentum for a source size of $r_0=1$ fm. We show the full TROY result with the ZR kernel (solid black line), the LL model (short-dashed red line), and the correlation function obtained with the on-shell version of the $T$-matrix (dash-dotted blue line).} \label{fig:D0barp_TROY}
\end{figure}

We start with the simple $\bar{D}^0 p$ correlation function in Fig.~\ref{fig:D0barp_TROY}. This is the unique channel in the $C=-1, S=0, Q=1$ sector. The TROY correlation function is very similar to the on-shell case and to the LL result, showing the effect of a repulsive interaction as expected from the negative value of the scattering length, $a_0=(-0.35 + \rm{i}0)$ fm.

\begin{figure}[h!]
\centering
\includegraphics[scale=0.55]{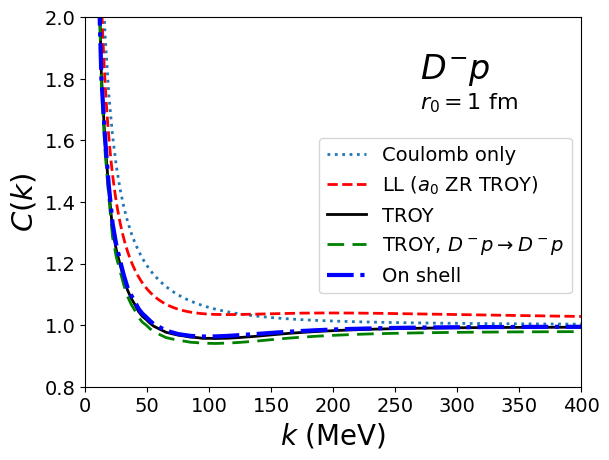}
\caption{$D^-p$ correlation function as function of the relative momentum for a source size of $r_0=1$ fm.  We show the full TROY result with the ZR kernel (solid black line), the elastic contribution to the TROY result (dashed green line), the LL model (short-dashed red line), and the correlation function obtained with the on-shell version of the $T$-matrix (dash-dotted blue line). The dotted blue line displays the Coulomb-only result. }  \label{fig:Dmp_TROY}
\end{figure}

In Fig.~\ref{fig:Dmp_TROY} we present the $D^-p$ correlation function. The full TROY result contains the effect of the $D^-p$ and $\bar{D}^0 n$ coupled channels present in this $C=-1, S=0, Q=0$ sector. The inelastic process $\bar{D}^0 n \rightarrow D^-p$ contributes with a thermal weight very close to one and its effect is extremely small, as can be seen by the difference between the solid black line and the elastic contribution represented by the long-dashed green line. The elastic channel has a scattering length of $a_0=(-0.16 + \rm{i}0.01)$ fm, signalling a repulsive strong interaction on top of the attractive Coulomb one, the correlation function of which is represented by the dotted blue line.
The comparison with the LL approximation shows some quantitative differences. While the TROY result remains always below the Coulomb only correlation function, and stays below one, the LL  result remains always above one, and crosses the Coulomb only correlation function around $k=120$ MeV.

\begin{figure}[h!]
\centering
\includegraphics[scale=0.55]{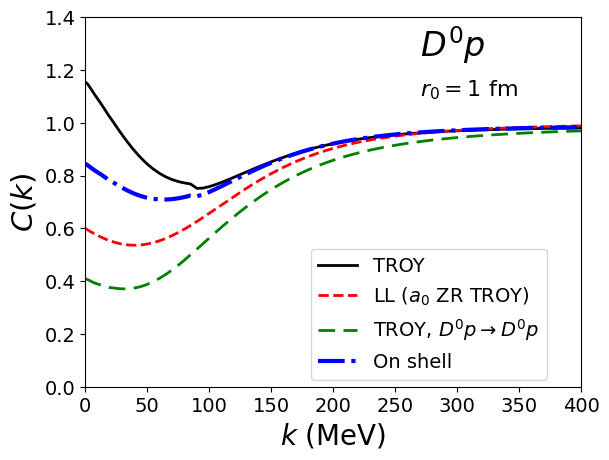}
\caption{$D ^0p$  correlation function as function of the relative momentum for a source size of $r_0=1$ fm. We show the full TROY result with the ZR kernel (solid black line), the elastic contribution to the TROY result (dashed green line), the LL model (short-dashed red line), and the correlation function obtained with the on-shell version of the $T$-matrix (dash-dotted blue line).  }  \label{fig:D0p_TROY}
\end{figure}

Moving to the charm one sector, we start by showing in Fig.~\ref{fig:D0p_TROY} the correlation function of $D^0 p$ pairs, free from the Coulomb interaction and involving 16 coupled  channels.
The full TROY result contains the contribution of all these channels with some thermal weights, giving a large contribution at low momentum values (4 channels are below the elastic one, and 11 channels above it, see Sec.~\ref{sec:coupled}).  The cusp observed at $k=88$ MeV signals the opening of the $D^+ n$ channel.
In this case the TROY and on-shell results differ more significantly, as the interaction is now sensitive to the existence of a resonance below but very close to threshold, the $\Sigma_c(2800)$ in $I=1$. This is reflected in the large negative real part of the $D^0 p$ scattering length, which is $a_0=(-1.62 + \rm{i}0.48)$ fm in the TROY formalism, to be compared with the on-shell value of $a_0=(-1.44 + \rm{i} 0.35)$ fm. Note that the $D^0 p$ interaction also develops a bound state with $I=0$, the $\Lambda_c(2595)$. This state should leave a visible effect in the correlation functions of $\Sigma_c \pi$ pairs, which are not the object of the present study.  
The elastic contribution, represented by the long-dashed green line, also differs considerably from the LL result, displayed by the short-dashed red line. This has to do with the short-distance behavior of the wave function, which differs significantly from the asymptotic form assumed by the LL approach. This is especially significant at low momentum, since the wave function is more affected by the influence of the subthreshold resonance.

\begin{figure}[h!]
\centering
\includegraphics[scale=0.55]{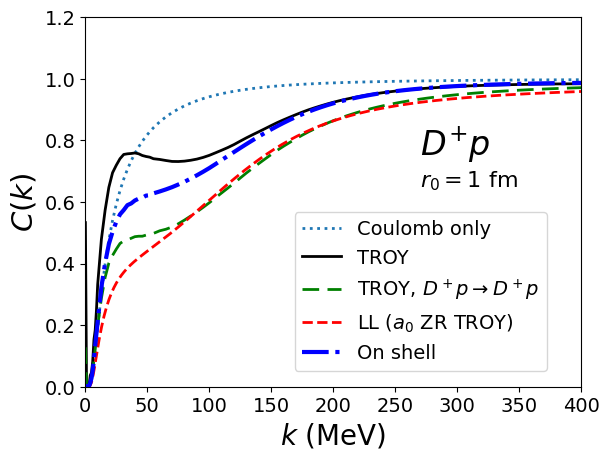}
\caption{$D^+p$  correlation function as function of the relative momentum for a source size of $r_0=1$ fm. We show the full TROY result with the ZR kernel (solid black line), the elastic contribution to the TROY result (dashed green line), the LL model (short-dashed red line), and the correlation function obtained with the on-shell version of the $T$-matrix (dash-dotted blue line). The dotted blue line displays the Coulomb-only result. }  \label{fig:Dpp_TROY}
\end{figure}

Finally, we focus on the $D^+p$ correlation as shown in Fig.~\ref{fig:Dpp_TROY}. This sector is affected by the Coulomb interaction and has 9 coupled channels. Again, the direct contribution to the correlation function is quite comparable to the LL result, except at low momentum as explained above.  Inelastic channels provide a sizable contribution, especially around $k=50$ MeV, as will be detailed in Sec.~\ref{sec:coupled}, probing the importance of coupled channels in the $D^+p$ case, where the LL approximation would not be appropriate.
The difference between the TROY and on-shell results can be understood by their different scattering lengths in the
elastic  $D^+p$ channel, namely $a_0=(-1.97 + \rm{i}0.39)$ fm and $a_0=(-1.79+ \rm{i}0.32)$ fm, respectively. As before, these sizable values are directly related to the $I=1$ $\Sigma_c(2800)$ resonance present in both models. 

In Table~\ref{tab:scat_lengths} we provide a summary of the $s$-wave scattering lengths for all channels, in both on-shell and off-shell approximations of the ZR interaction.

\begin{widetext}

\begin{table}[h!]
\centering
\begin{tabular}{cccc}
\hline \hline
Channel & $a_0$ (fm) ZR TROY & $a_0$ (fm) ZR on shell (charge) & $a_0$ (fm) ZR on shell (isospin) \\ \hline
$\bar{D}^0 p$ & $-0.35+\rm{i}0$ & $-0.34+\rm{i} 0 $ & $-0.34+\rm{i} 0 $ \\
$D^- p$ & $-0.16+\rm{i}0.01$ & $-0.16+\rm{i} 0.01 $ &$-0.16+\rm{i} 0 $ \\
$D^0 p$ & $-1.62+\rm{i}0.48$ &$-1.44+\rm{i} 0.35 $ &$-1.21+\rm{i} 0.16 $ \\
$D^+ p$ & $-1.97+\rm{i}0.39$ &$-1.79+\rm{i} 0.32 $ & $-1.79+\rm{i} 0.32 $ \\
 \hline \hline
\end{tabular}
\caption{$s-$wave scattering length for $DN$ and $\bar{D}N$ channels in the ZR approximation using the off-shell unitarization implemented in TROY, and the on-shell factorization.}
\label{tab:scat_lengths}
\end{table}
\end{widetext}

Before concluding this section, we would like to comment on the concerns raised in Ref.~\cite{Epelbaum:2025aan} about the use of a universal source function and the resulting ambiguities in the correlation function tied to the off-shell behavior of the wave-function or, equivalently, to the off-shell dependence of the $T$ matrix. 
We indeed find that the way the off-shell treatment is done, such as the type of form-factor employed, or the value of the cut-off, influences the results, especially in the $DN$ sector. However, using physically motivated parameters, fine tuned to experimental constraints (like the position of the generated resonances) help to minimize this dependence. In fact, the interactions used in the present work can be thought of being built by the exchange of vector mesons and a cut-off of the order of the vector mass ($600-1000$~MeV), as selected here, should be appropriate. Besides, the study on Ref.~\cite{Molina:2025lzw} explicitly shows, for a meson-baryon interaction of the type of the present work but in the strangeness $S=0$ sector around the $N^*(1535)$ resonance, that the uncertainties in the correlation functions due to off-shell ambiguities are of the order of 2–3\%.

\subsection{Coupled channels in the $D^+p$ and $D^0p$ correlation functions~\label{sec:coupled}}

In this section, we comment on the different channel contributions to the correlation function of $DN$ pairs. We first recall that the thermal weight $w_i$ associated to each channel coupling to the measured $DN$ pair has been obtained following the prescription of Eqs.~(\ref{eq:thermal_mult}) and (\ref{eq:weight}). 
For further reference, the resulting values are provided in Table~\ref{tab:weights} for $D^+p$ pairs (left side) and $D^0p$ pairs (right side) and their related coupled channels.

\begin{table}[h!]
\centering
\begin{tabular}{cc|cc}
\hline \hline
Pair, $i$ & $w_i$ & Pair, $i$ & $w_i$ \\ \hline
$\pi^+ \Lambda_c^+$ & 2.20 & $\pi^0 \Lambda_c^+$  & 2.16 \\
$\pi^0 \Sigma_c^{++}$ & 0.92 & $\pi^0 \Sigma_c^{+}$ & 0.90 \\
$\pi^+ \Sigma_c^+$ & 0.91 & $\pi^+ \Sigma_c^{0}$ & 0.90 \\
\bm{$D^+ p$} & {\bf 1.00} & $\pi^- \Sigma_c^{++}$ & 0.90 \\
$K^+ \Xi_c^+$ & 0.29 & \bm{$D^0p$} & {\bf 1.00} \\
$\eta \Sigma_c^{++}$ & 0.26 & $D^+n$ & 0.97 \\
$K^{+} \Xi'^+_c$ & 0.16 & $\eta \Lambda_c^+$ &  0.60 \\
$D_s^+ \Sigma^+$ & 0.19 & $K^+ \Xi_c^0$ & 0.28 \\
$\eta' \Sigma_c^{++}$ & 0.04 & $K^0 \Xi_c^+$ & 0.28 \\
& &  $\eta \Sigma_c^{+}$ & 0.25 \\
& &  $K^+ \Xi_c'^0$ & 0.16 \\
& &  $K^0 \Xi_c'^+$ & 0.16 \\
& &  $D_s^+ \Lambda^0$ & 0.26 \\
& &  $D_s^+ \Sigma^0$ &  0.18 \\
& &  $\eta' \Lambda_c^+$ & 0.10 \\
& &  $\eta' \Sigma_c^+$ &  0.04 \\
 \hline \hline
\end{tabular}
\caption{Thermal weights, as computed by the prescription described in Sec.~\ref{sec:TROY}, of the $D^+p$ (left side) and $D^0p$ (right side) cases and their coupled channels.}
\label{tab:weights}
\end{table}

\begin{widetext}
    
\begin{figure}[h!]
\centering
\includegraphics[scale=0.55]{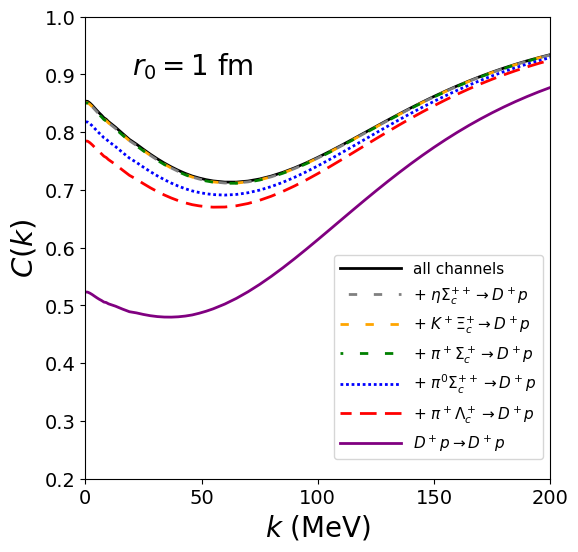}
\includegraphics[scale=0.55]{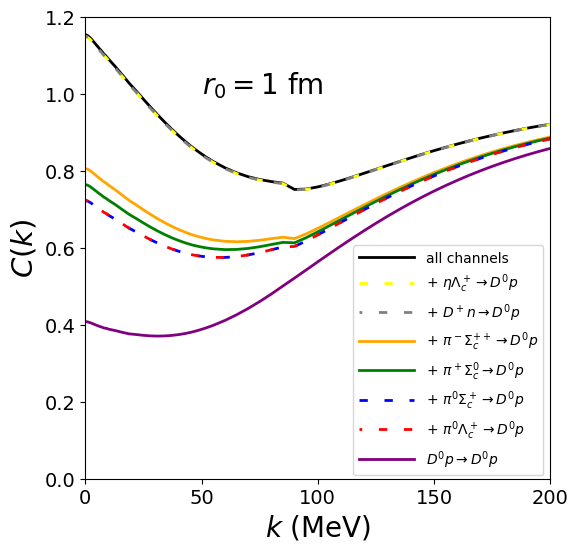}
\caption{Correlation function of $DN$ for a Gaussian source with $r_0=1$ fm. Left panel: $D^+p$ with only strong interaction. Right panel: $D^0p$. The lines are to be read from bottom to the top when we add additional channels up to complete all of them (``all channels'')~\label{fig:decomp}}\end{figure}

\end{widetext}

The decomposition of the $D^+p$ and the $D^0p$ correlation functions for a source size of $r_0=1$ fm is shown in Fig.~\ref{fig:decomp}. 

The left panel of this figure focuses on the $D^+p$ correlation function. The main contribution comes from the elastic channel, but the remaining ones (added one by one from the bottom to the top) contribute considerably. Notice that this channel is only affected by the $I=1$ sector of the $DN$ interaction, where the dynamically generated $\Sigma_c (2800)$ appears, a state that has also been experimentally seen in $\pi\Lambda_c^+$ spectra \cite{ParticleDataGroup:2024cfk}. According to Ref.~\cite{Jimenez-Tejero:2011dif} this resonance couples very strongly to $DN$. It also has a non-negligible coupling to the $\pi \Lambda_c$ and $\pi \Sigma_c$ channels, the size of the former being twice that of the latter. These two channels also have the largest thermal weights, since their thresholds are located below the $DN$ one. The combined effect of the weight and the coupling to the $\Sigma_c(2800)$ explains the large contribution of the $\pi \Lambda_c$ channel (difference between the solid purple and dashed red lines), followed by that of the $\pi \Sigma_c$ channels (difference between the dotted blue and dashed red lines; and between the dashed green and dotted blue lines). The rest of the channels have smaller weights, so their effect is negligible. We conclude that a measurement of the $D^+p$ correlation function would be very sensitive to the presence of the $\Sigma_c(2800)$ and would bring very valuable information to constrain the models of the $DN$ interaction.

In the right panel, we plot the $D^0p$ case and its coupled channels, one by one, from the bottom to the top. Here we observe a large effect of the $\pi \Lambda_c$ channel, for the same reason as before, that is, because of its sizable coupling to the $I=1$ $\Sigma_c(2800)$ resonance and the larger value of the thermal weight. For the $\pi \Sigma_c$ case we observe a moderate effect from the $\pi^+ \Sigma_c^0$ (difference between the solid green and dashed blue lines) and $\pi^- \Sigma_c^{++}$ (difference between the solid orange and solid green lines) channels, which are coupled to the $D^0p$ state in both the $I=0$ and $I=1$ channels, dominated respectively, by the $\Lambda_c(2595)$ and the $\Sigma_c(2800)$ resonances. Interestingly, the effect of the $\pi^0 \Sigma_c^+$ (difference between the dashed red and blue lines) is almost zero. We note that this state only couples to the $D^0p$ via the $I=0$ channel (the $I=1$ has zero Clebsch-Gordan coefficient). We therefore conclude that the relevance of a channel in the $D^0p$ correlation function is driven by its coupling to the $\Sigma_c(2800)$ resonance, which lies close to the kinematical region of interest. Finally, we observe that the largest coupled channel contribution is due to the $D^+n$ channel, which has a threshold very close to the $D^0 p$ one, and therefore a similar thermal weight. This contribution is enhanced by the fact that the $DN$ coupling to the  $\Sigma_c(2800)$ resonance is about an order of magnitude larger than that of the $\pi\Lambda_c$ and $\pi\Sigma_c$ states \cite{Jimenez-Tejero:2011dif}. The channels with thresholds above the $D^+n$ one provide a negligible contribution to the $D^0 p$ correlation function.

\subsection{Dependence on the source size and comparison with data~\label{sec:size}}

\begin{figure}[h!]
\centering
\includegraphics[scale=0.55]{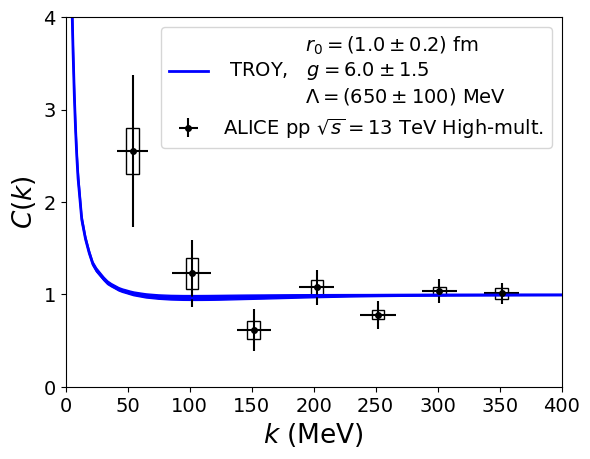}
\caption{$D^-p$ correlation function using TROY solution of the ZR approximation. The Gaussian 
 source radius is varied between 0.8 fm and 1.2 fm, relevant for pp collisions, the coupling constant $g$ between 4.5 and 7.5, and the hard cut-off $\Lambda$ between 550 MeV and 750 MeV
radii is varied between $r_0=(0.9 \pm 0.1)$ fm.} \label{fig:Dmp_ALICE}
\end{figure}

In Fig.~\ref{fig:Dmp_ALICE} we plot the $D^-p$ correlation function using TROY for the meson-baryon interaction in the ZR approximation. It represents an overall (Coulomb plus strong)  attractive interaction, which cannot fully explain the current experimental data by ALICE~\cite{ALICE:2022enj}. We note that the blue band displaying our results collects the correlation functions of 2000 configurations obtained by varying the Gaussian source radius between 0.8 fm and 1.2 fm, relevant for pp collisions, the coupling constant $g$ between 4.5 and 7.5, and the hard cut-off $\Lambda$ between 550 MeV and 750 MeV. Even with that flexibility in the model parameters we are unable to reproduce the current ALICE data. In Ref.~\cite{ALICE:2022enj} only the parametrizations by Haidenbauer---with an attractive interaction in the strong sector given by an additional strength of the scalar exchange---and Yamaguchi, with a bound state, show a reasonable agreement with the data. In the Yamaguchi case, as can be seen in our Fig.~\ref{fig:CLLD-Comparacion}, we obtain a result consistent with the presence of the bound state. 

\begin{figure}[h!]
\centering
\includegraphics[scale=0.55]{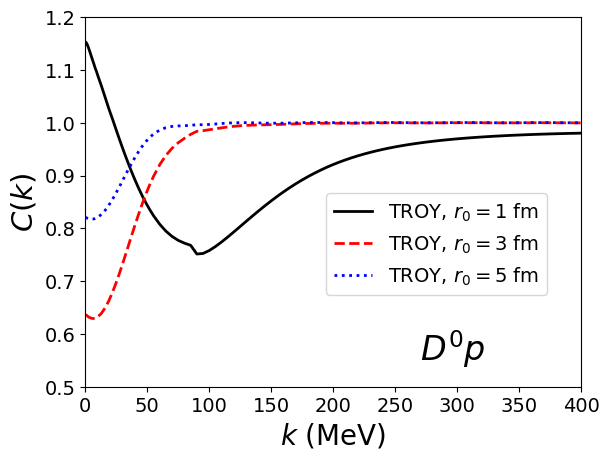}
\includegraphics[scale=0.55]{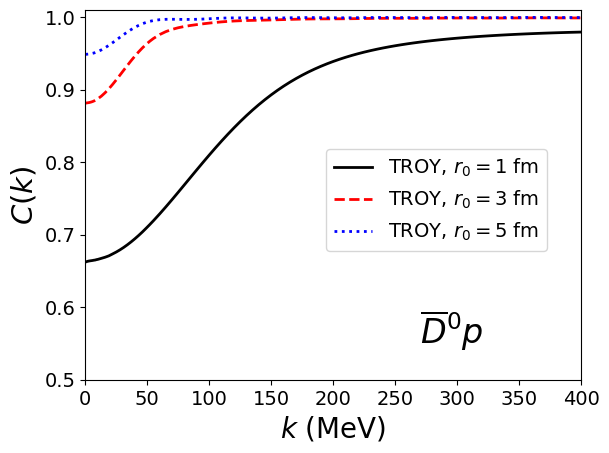}
\caption{$D^0p$ (top) and $\bar{D}^0p$ (bottom) correlation functions using TROY solution of the ZR approximation with thermal weights. The Gaussian radii are taken to be 1 fm, 3 fm, and 5 fm to cover larger systems as those measured by the STAR collaboration in these sectors}  \label{fig:D0D0barp_TROY_R}
\end{figure}

In Fig.~\ref{fig:D0D0barp_TROY_R} we present the results of the $D^0p$ and $\bar{D}^0p$ channels using the full coupled-channel TROY formalism with thermal weights. The Gaussian radii  are chosen as $r_0=1$ fm, 3 fm, and 5 fm. The latter is used to mimic the source sizes expected in Au+Au collisions measured by the STAR collaboration. The signal for these large radii is rather weak and suppressed, but our result is roughly consistent with the preliminary data shown in Ref.~\cite{RoyChowdhury:2025xhv} for the larger radii. However, in that reference, the $D^0p$ and $\bar{D}^0p$ cases are combined into a single correlation function, even if they are not charge conjugate channels.
Since the interaction is rather different for each of these sectors, we doubt that this is a sensible comparison and urge our experimental colleagues to provide each channel separately.

\section{Summary~\label{sec:Conclusions}}

In this work we have studied the femtoscopic correlation function of $D^+p, D^0p, D^-p$ and $\bar{D}^0p$ pairs applying the Koonin-Pratt equation. The meson-baryon interaction has been extracted from an effective field theory at low energy, working in the zero-range approximation. Non perturbative $T$-matrix equations have been implemented both in the on-shell factorization approach, as well as in the off-shell case using the TROY formalism. These two methods provide different approximations for the $l=0$ wave function of the meson-baryon pair, which eventually are used in the Koonin-Pratt formula. Coulomb effects have also been incorporated in the appropriate channels. 
Our results have been compared to those obtained within the Lednick\'y-Lyuboshitz approximation of the correlation function. We remind the reader that this approximation is based on the asymptotic behavior of the wave function of the direct channel, thereby neglecting the effect of the inelastic components as well as possible deviations of the wave function at small distances, which can be important at low momentum if a nearby subthreshold resonance is present. For the $D^+p$ case we found that,(with the exception of the SU(8) calculation, different results in the literature provide a similar correlation function reflecting an apparently repulsive strong interaction. For the $D^-p$ pair we also find consistency among the different results despite the different values of the scattering lengths, with a clear difference of the Haidenbauer's calculation including an enhanced $\sigma$ exchange which encodes a rather strong attraction. This result was close to the experimental data as shown in Ref.~\cite{ALICE:2018nnl}. We stress that in our version of the LL formula, the Yamaguchi result does not approach the data, but shows the typical effect of a bound state, in accordance with the theoretical expectations.

The full TROY calculation incorporates all the inelastic channels and obtains the complete spatial dependence of the wave function (not just the asymptotic part) out of the off-shell $T$-matrix elements. Moreover, we have estimated the thermal weights for all channels, taking into account the masses of the produced pairs in a system at finite temperature for the conditions of high-multiplicity pp collisions at $\sqrt{s}=13$ TeV in the LHC.
For the $D^+p$ case we found the direct elastic channel of our model to be roughly compatible with the LL approximation, but the effect of the coupled channels is non negligible, increasing the correlation function between 30-50\% at low momentum. We provide this correlation function for a small source radius as a prediction for future measurements of this system in the ALICE experiment, in which we foresee that an analysis based on the LL approach will not be appropriate enough. For the $D^-p$ case, the effect of the only coupled channel $(\bar{D}^0 n)$ is found to be minimal with respect to the direct one, but  there are differences between the LL and TROY calculations of the correlation function, the latter maintaining values above one for all momenta, while the former approaches one from below. We find no consistency with the experimental data for $D^+p$ pairs measured by ALICE in pp collisions. Finally, we have computed the correlation function for neutral channels, $D^0p$ and $\bar{D}^0p$, motivated by the preliminary results by the STAR collaboration. Since these correspond to Au+Au collisions we scanned large Gaussian radii between $r_0=1$ fm and $r_0=5$ fm. We have not quantitatively compared with preliminary data since they are given as the sum of the two channels, and we cannot really assess the different nature of the interaction in the $C=1$ and $C=-1$ sectors.

To further test the results of this paper, it would be vital to compare with the $D^+p$ correlation function in small systems like pp collisions, for which we present a prediction, but no experimental results have been yet provided by the ALICE experiment. This channel could be an excellent way to study the effect of the $\Sigma_c (2800)$ resonance. In addition, it would be interesting to compare with STAR results in larger systems, like the one shown in~\cite{RoyChowdhury:2025xhv}, but after the data of the two charm sectors are given separately. While this would lead to an increase of the statistical uncertainties, it would make the comparison more sensible, since in our model the $D^0p$ and $\bar{D}^0p$ interactions are rather different, namely, strongly attractive and repulsive, respectively.

\begin{acknowledgments}
We acknowledge support from the following national agencies: Unidades de Excelencia María de Maeztu CEX2020-001058-M and CEX2024-001451-M, and Projects No. PID2022-139427NB-I00 and No.PID2023-147112NB-C21, funded by MICIU/AEI/10.13039/501100011033 (Spain); Contract 2021 SGR 171 by the Generalitat de Catalunya; CRC-TR 211 ’Strong-interaction matter under extreme conditions’, Project No. 315477589 - TRR 211 by the Deutsche Forschungsgemeinshaft;  Grant No. 402942/2024 by the Brazilian CNPq (National Council for Scientific and Technological
Development) and Grant CIPROM 2023/59 of Generalitat Valenciana.

\end{acknowledgments}


\appendix

\section{The Coulomb interaction}\label{ap:coulomb}

In order to account for channels involving a pair of charged particles, one needs to include the Coulomb force. As discussed in our previous work~\cite{Torres-Rincon:2024znb}, our approach consists in adding  the effect of the Coulomb interaction in momentum space, inspired by Refs.~\cite{joachain1975quantum,Holzenkamp:1989tq}, to the tree-level strong interaction potential, that is,  Eq.~(\ref{eq:VWT}) for the WT interaction and Eq.~(\ref{eq:VZR}) for the ZR approximation.

In momentum space the Coulomb interaction is obtained by Fourier transforming the potential in coordinate space,
\begin{eqnarray}
 &&V^{\rm C}(| \boldsymbol{p}^\prime- \boldsymbol{p} |; \mathcal{ R}_C) = \nonumber \\ && \int^{\mathcal{ R}_C}_{0} d^3r \ {\rm e}^{ i (\boldsymbol{p}^\prime-\boldsymbol{p}) \cdot \boldsymbol{r} } \ \frac{\varepsilon\alpha}{r} = \frac{4\pi \varepsilon \alpha}{| \boldsymbol{p}^\prime- \boldsymbol{p}  |^2}\left[ 1-\cos( |\boldsymbol{p}^\prime- \boldsymbol{p}  | \mathcal{ R}_C)\right] \ , \nonumber \\
 \label{eq:coul}
\end{eqnarray} 
where $r <  \mathcal{ R}_C$ so as to regulate its long-range character. In this way, our approach becomes numerically tractable avoiding the forward scattering singularity of the Coulomb interaction at $|\boldsymbol{p}'-\boldsymbol{p}|=0$. 

The $l=0$ component of the potential in Eq.~(\ref{eq:coul}) is given by
\begin{align}
&V^{\rm C}_{\rm s-wave}(p,p^\prime; \mathcal{ R}_C) \nonumber \\ &=  \frac{1}{2} \int_{-1}^{1} d\cos \theta_{\boldsymbol{pp'}} \ V^{\rm C}(|\boldsymbol{p}^\prime- \boldsymbol{p} |; \mathcal{ R}_C) \nonumber \\
&= \frac{2\pi \varepsilon \alpha}{p p^\prime} \left\{ {\rm Ci} \left[ |p^\prime - p| \mathcal{ R}_C \right] - {\rm Ci}\left[ (p^\prime +p) \mathcal{ R}_C \right] + \ln\left(\frac{p^\prime + p}{| p^\prime - p| }\right)\right\}  \ ,
 \label{eq:s-wave_coul}
\end{align} 
where ${\rm Ci}[x]=\int_x^\infty dt \ (\cos t)/t$ is the cosine integral function. In the following, we omit the label $\mathcal{ R}_C$ to simplify the notation.

In order to add the projected Coulomb interaction to the strong one, two modifications are needed, which are related to the nonrelativistic character of the Coulomb force, as the scattering amplitude $T$ comes from the
relativistic Bethe-Salpeter equation (see Eqs.~(\ref{eq:os},\ref{eq:gloop})). 

On the one hand, given that the two-body propagator employs relativistic energies,  the following replacement is needed
\be  V^{\rm C}_{\rm s-wave}(p,p^\prime) \longrightarrow
\sqrt{\xi(p;s)} \ V^{\rm C}_{\rm s-wave}(p,p^\prime) \ \sqrt{\xi (p^\prime;s)} \ ,  \ee
where the kinematic factors $\xi$ are given by

\be \xi(p;s)= 2 \mu \frac{{s}-(E_1(p)+\omega_2(p))^2}{2(E_1(p)+\omega_2(p))} \frac{1}{ \frac{\lambda(s,M_1,m_2)}{4s}-p^2} \ ,
\ee 
with $\mu=M_1 m_2 /(M_1+m_2)$ being the reduced mass of a baryon with mass $M_1$ and energy $E_1(p)$, and a meson with mass $m_2$ and energy $\omega_2(p)$. The quantity  $\lambda(s,M_1,m_2)=[(s-(M_1+m_2)^2][s-(M_1-m_2)^2]$ is the K\"all\'en function.

On the other hand, we have to compensate for the normalization factors of type $1/\displaystyle\sqrt{2\omega}$ and $2M/2E$ that are linked to the relativistic treatment of the two-body propagator in Eqs.~(\ref{eq:os},\ref{eq:gloop}), which do  not appear in a Lippmann-Schwinger-type formulation appropriate for the non-relativistic Coulomb interaction. 

Hence, the  $s-$wave Coulomb contribution to be added to  the strong interaction kernel (only in diagonal transitions involving a pair of charged particles) reads
\begin{widetext}
\begin{align} 
V^{\rm C,rel}_{\rm s-wave}(p,p^\prime;\sqrt{s}) & = \sqrt{\frac{2E_1(p)}{2M_1}} \sqrt{2\omega_2(p)} \sqrt{\xi(p;s)}  V^{\rm C}_{\rm s-wave} (p,p^\prime)  \sqrt{\frac{2E_1(p^\prime)}{2M_1}} \sqrt{2\omega_2(p^\prime)} \sqrt{\xi(p^\prime;s)}  \ . 
\label{eq:vc_rel}
\end{align}
\end{widetext}

\begin{widetext}

\begin{figure}[h!]
\centering
\includegraphics[width=0.45\textwidth]{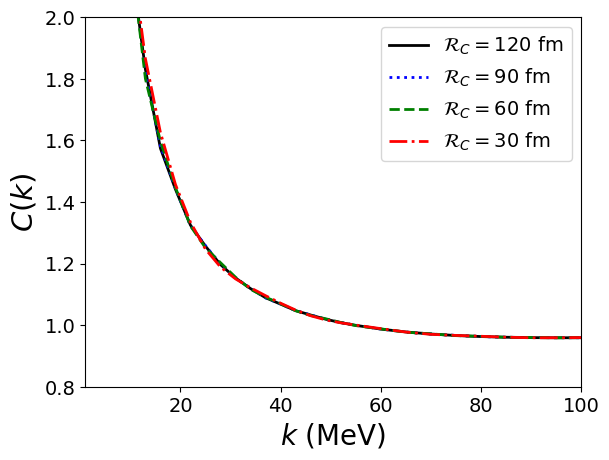}
\includegraphics[width=0.45\textwidth]{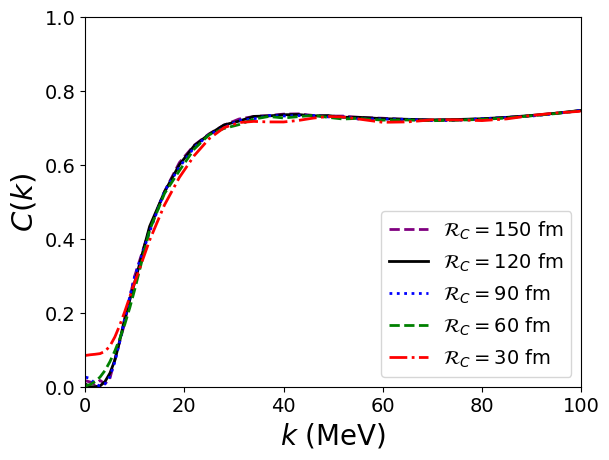}
\caption{Dependence of the correlation functions for $D^-p$ (left) and $D^+p$ (right) on the $\mathcal{R }_C$ coefficient. Both correlation functions are calculated with TROY in the ZR approximation of the interaction. The Gaussian source radius is chosen to be $r_0=1$ fm.}  \label{fig:RCdependence}
\end{figure}

\end{widetext}

Finally, we comment on the dependence of the correlation functions with $\mathcal{ R}_C$. This parameter cannot be asymptotically large due to the forward scattering singularity, but also not too small since it should reflect the long range nature of the interaction. We find that for the $D^-p$ system, the resulting $C(k)$ converges already for several dozens of fm, and in the conservative side we have set $\mathcal{ R}_C=60$ fm. The left panel of Fig.~\ref{fig:RCdependence} shows the dependence of this correlation function with TROY for a source radius of $r_0=1$ fm. For the $D^+p$ case, we find a much higher sensitivity to this parameter. As can be seen in the right panel of Fig.~\ref{fig:RCdependence}, a value of $\mathcal{ R}_C=30$ fm  still presents convergence problems at small momenta. These effects disappear as $\mathcal{ R}_C$ increases. In the results of this paper, we have used a conservative value of $\mathcal{ R}_C=120$ fm for this system.

\bibliography{refs}

\begin{thebibliography}{69}%
\makeatletter
\providecommand \@ifxundefined [1]{%
 \@ifx{#1\undefined}
}%
\providecommand \@ifnum [1]{%
 \ifnum #1\expandafter \@firstoftwo
 \else \expandafter \@secondoftwo
 \fi
}%
\providecommand \@ifx [1]{%
 \ifx #1\expandafter \@firstoftwo
 \else \expandafter \@secondoftwo
 \fi
}%
\providecommand \natexlab [1]{#1}%
\providecommand \enquote  [1]{``#1''}%
\providecommand \bibnamefont  [1]{#1}%
\providecommand \bibfnamefont [1]{#1}%
\providecommand \citenamefont [1]{#1}%
\providecommand \href@noop [0]{\@secondoftwo}%
\providecommand \href [0]{\begingroup \@sanitize@url \@href}%
\providecommand \@href[1]{\@@startlink{#1}\@@href}%
\providecommand \@@href[1]{\endgroup#1\@@endlink}%
\providecommand \@sanitize@url [0]{\catcode `\\12\catcode `\$12\catcode
  `\&12\catcode `\#12\catcode `\^12\catcode `\_12\catcode `\%12\relax}%
\providecommand \@@startlink[1]{}%
\providecommand \@@endlink[0]{}%
\providecommand \url  [0]{\begingroup\@sanitize@url \@url }%
\providecommand \@url [1]{\endgroup\@href {#1}{\urlprefix }}%
\providecommand \urlprefix  [0]{URL }%
\providecommand \Eprint [0]{\href }%
\providecommand \doibase [0]{http://dx.doi.org/}%
\providecommand \selectlanguage [0]{\@gobble}%
\providecommand \bibinfo  [0]{\@secondoftwo}%
\providecommand \bibfield  [0]{\@secondoftwo}%
\providecommand \translation [1]{[#1]}%
\providecommand \BibitemOpen [0]{}%
\providecommand \bibitemStop [0]{}%
\providecommand \bibitemNoStop [0]{.\EOS\space}%
\providecommand \EOS [0]{\spacefactor3000\relax}%
\providecommand \BibitemShut  [1]{\csname bibitem#1\endcsname}%
\let\auto@bib@innerbib\@empty
\bibitem [{\citenamefont {Fabbietti}\ \emph {et~al.}(2021)\citenamefont
  {Fabbietti}, \citenamefont {Mantovani~Sarti},\ and\ \citenamefont
  {Vazquez~Doce}}]{Fabbietti:2020bfg}%
  \BibitemOpen
  \bibfield  {author} {\bibinfo {author} {\bibfnamefont {L.}~\bibnamefont
  {Fabbietti}}, \bibinfo {author} {\bibfnamefont {V.}~\bibnamefont
  {Mantovani~Sarti}}, \ and\ \bibinfo {author} {\bibfnamefont {O.}~\bibnamefont
  {Vazquez~Doce}},\ }\href {\doibase 10.1146/annurev-nucl-102419-034438}
  {\bibfield  {journal} {\bibinfo  {journal} {Ann. Rev. Nucl. Part. Sci.}\
  }\textbf {\bibinfo {volume} {71}},\ \bibinfo {pages} {377} (\bibinfo {year}
  {2021})},\ \Eprint {http://arxiv.org/abs/2012.09806} {arXiv:2012.09806
  [nucl-ex]} \BibitemShut {NoStop}%
\bibitem [{\citenamefont {Adamczewski-Musch}\ \emph {et~al.}(2016)\citenamefont
  {Adamczewski-Musch} \emph {et~al.}}]{HADES:2016dyd}%
  \BibitemOpen
  \bibfield  {author} {\bibinfo {author} {\bibfnamefont {J.}~\bibnamefont
  {Adamczewski-Musch}} \emph {et~al.} (\bibinfo {collaboration} {HADES}),\
  }\href {\doibase 10.1103/PhysRevC.94.025201} {\bibfield  {journal} {\bibinfo
  {journal} {Phys. Rev. C}\ }\textbf {\bibinfo {volume} {94}},\ \bibinfo
  {pages} {025201} (\bibinfo {year} {2016})},\ \Eprint
  {http://arxiv.org/abs/1602.08880} {arXiv:1602.08880 [nucl-ex]} \BibitemShut
  {NoStop}%
\bibitem [{\citenamefont {Shapoval}\ \emph {et~al.}(2015)\citenamefont
  {Shapoval}, \citenamefont {Erazmus}, \citenamefont {Lednicky},\ and\
  \citenamefont {Sinyukov}}]{Shapoval:2014yha}%
  \BibitemOpen
  \bibfield  {author} {\bibinfo {author} {\bibfnamefont {V.~M.}\ \bibnamefont
  {Shapoval}}, \bibinfo {author} {\bibfnamefont {B.}~\bibnamefont {Erazmus}},
  \bibinfo {author} {\bibfnamefont {R.}~\bibnamefont {Lednicky}}, \ and\
  \bibinfo {author} {\bibfnamefont {Y.~M.}\ \bibnamefont {Sinyukov}},\ }\href
  {\doibase 10.1103/PhysRevC.92.034910} {\bibfield  {journal} {\bibinfo
  {journal} {Phys. Rev. C}\ }\textbf {\bibinfo {volume} {92}},\ \bibinfo
  {pages} {034910} (\bibinfo {year} {2015})},\ \Eprint
  {http://arxiv.org/abs/1405.3594} {arXiv:1405.3594 [nucl-th]} \BibitemShut
  {NoStop}%
\bibitem [{\citenamefont {Adamczyk}\ \emph
  {et~al.}(2015{\natexlab{a}})\citenamefont {Adamczyk} \emph
  {et~al.}}]{STAR:2014dcy}%
  \BibitemOpen
  \bibfield  {author} {\bibinfo {author} {\bibfnamefont {L.}~\bibnamefont
  {Adamczyk}} \emph {et~al.} (\bibinfo {collaboration} {STAR}),\ }\href
  {\doibase 10.1103/PhysRevLett.114.022301} {\bibfield  {journal} {\bibinfo
  {journal} {Phys. Rev. Lett.}\ }\textbf {\bibinfo {volume} {114}},\ \bibinfo
  {pages} {022301} (\bibinfo {year} {2015}{\natexlab{a}})},\ \Eprint
  {http://arxiv.org/abs/1408.4360} {arXiv:1408.4360 [nucl-ex]} \BibitemShut
  {NoStop}%
\bibitem [{\citenamefont {Adamczyk}\ \emph
  {et~al.}(2015{\natexlab{b}})\citenamefont {Adamczyk} \emph
  {et~al.}}]{STAR:2015kha}%
  \BibitemOpen
  \bibfield  {author} {\bibinfo {author} {\bibfnamefont {L.}~\bibnamefont
  {Adamczyk}} \emph {et~al.} (\bibinfo {collaboration} {STAR}),\ }\href
  {\doibase 10.1038/nature15724} {\bibfield  {journal} {\bibinfo  {journal}
  {Nature}\ }\textbf {\bibinfo {volume} {527}},\ \bibinfo {pages} {345}
  (\bibinfo {year} {2015}{\natexlab{b}})},\ \Eprint
  {http://arxiv.org/abs/1507.07158} {arXiv:1507.07158 [nucl-ex]} \BibitemShut
  {NoStop}%
\bibitem [{\citenamefont {Adam}\ \emph {et~al.}(2019)\citenamefont {Adam} \emph
  {et~al.}}]{STAR:2018uho}%
  \BibitemOpen
  \bibfield  {author} {\bibinfo {author} {\bibfnamefont {J.}~\bibnamefont
  {Adam}} \emph {et~al.} (\bibinfo {collaboration} {STAR}),\ }\href {\doibase
  10.1016/j.physletb.2019.01.055} {\bibfield  {journal} {\bibinfo  {journal}
  {Phys. Lett. B}\ }\textbf {\bibinfo {volume} {790}},\ \bibinfo {pages} {490}
  (\bibinfo {year} {2019})},\ \Eprint {http://arxiv.org/abs/1808.02511}
  {arXiv:1808.02511 [hep-ex]} \BibitemShut {NoStop}%
\bibitem [{\citenamefont {Acharya}\ \emph
  {et~al.}(2017{\natexlab{a}})\citenamefont {Acharya} \emph
  {et~al.}}]{ALICE:2017jto}%
  \BibitemOpen
  \bibfield  {author} {\bibinfo {author} {\bibfnamefont {S.}~\bibnamefont
  {Acharya}} \emph {et~al.} (\bibinfo {collaboration} {ALICE}),\ }\href
  {\doibase 10.1016/j.physletb.2017.09.009} {\bibfield  {journal} {\bibinfo
  {journal} {Phys. Lett. B}\ }\textbf {\bibinfo {volume} {774}},\ \bibinfo
  {pages} {64} (\bibinfo {year} {2017}{\natexlab{a}})},\ \Eprint
  {http://arxiv.org/abs/1705.04929} {arXiv:1705.04929 [nucl-ex]} \BibitemShut
  {NoStop}%
\bibitem [{\citenamefont {Acharya}\ \emph
  {et~al.}(2017{\natexlab{b}})\citenamefont {Acharya} \emph
  {et~al.}}]{ALICE:2017iga}%
  \BibitemOpen
  \bibfield  {author} {\bibinfo {author} {\bibfnamefont {S.}~\bibnamefont
  {Acharya}} \emph {et~al.} (\bibinfo {collaboration} {ALICE}),\ }\href
  {\doibase 10.1103/PhysRevC.96.064613} {\bibfield  {journal} {\bibinfo
  {journal} {Phys. Rev. C}\ }\textbf {\bibinfo {volume} {96}},\ \bibinfo
  {pages} {064613} (\bibinfo {year} {2017}{\natexlab{b}})},\ \Eprint
  {http://arxiv.org/abs/1709.01731} {arXiv:1709.01731 [nucl-ex]} \BibitemShut
  {NoStop}%
\bibitem [{\citenamefont {Acharya}\ \emph
  {et~al.}(2019{\natexlab{a}})\citenamefont {Acharya} \emph
  {et~al.}}]{ALICE:2018nnl}%
  \BibitemOpen
  \bibfield  {author} {\bibinfo {author} {\bibfnamefont {S.}~\bibnamefont
  {Acharya}} \emph {et~al.} (\bibinfo {collaboration} {ALICE}),\ }\href
  {\doibase 10.1016/j.physletb.2018.12.033} {\bibfield  {journal} {\bibinfo
  {journal} {Phys. Lett. B}\ }\textbf {\bibinfo {volume} {790}},\ \bibinfo
  {pages} {22} (\bibinfo {year} {2019}{\natexlab{a}})},\ \Eprint
  {http://arxiv.org/abs/1809.07899} {arXiv:1809.07899 [nucl-ex]} \BibitemShut
  {NoStop}%
\bibitem [{\citenamefont {Acharya}\ \emph
  {et~al.}(2019{\natexlab{b}})\citenamefont {Acharya} \emph
  {et~al.}}]{ALICE:2018ysd}%
  \BibitemOpen
  \bibfield  {author} {\bibinfo {author} {\bibfnamefont {S.}~\bibnamefont
  {Acharya}} \emph {et~al.} (\bibinfo {collaboration} {ALICE}),\ }\href
  {\doibase 10.1103/PhysRevC.99.024001} {\bibfield  {journal} {\bibinfo
  {journal} {Phys. Rev. C}\ }\textbf {\bibinfo {volume} {99}},\ \bibinfo
  {pages} {024001} (\bibinfo {year} {2019}{\natexlab{b}})},\ \Eprint
  {http://arxiv.org/abs/1805.12455} {arXiv:1805.12455 [nucl-ex]} \BibitemShut
  {NoStop}%
\bibitem [{\citenamefont {Acharya}\ \emph
  {et~al.}(2020{\natexlab{a}})\citenamefont {Acharya} \emph
  {et~al.}}]{ALICE:2019igo}%
  \BibitemOpen
  \bibfield  {author} {\bibinfo {author} {\bibfnamefont {S.}~\bibnamefont
  {Acharya}} \emph {et~al.} (\bibinfo {collaboration} {ALICE}),\ }\href
  {\doibase 10.1016/j.physletb.2020.135223} {\bibfield  {journal} {\bibinfo
  {journal} {Phys. Lett. B}\ }\textbf {\bibinfo {volume} {802}},\ \bibinfo
  {pages} {135223} (\bibinfo {year} {2020}{\natexlab{a}})},\ \Eprint
  {http://arxiv.org/abs/1903.06149} {arXiv:1903.06149 [nucl-ex]} \BibitemShut
  {NoStop}%
\bibitem [{\citenamefont {Acharya}\ \emph
  {et~al.}(2019{\natexlab{c}})\citenamefont {Acharya} \emph
  {et~al.}}]{ALICE:2019eol}%
  \BibitemOpen
  \bibfield  {author} {\bibinfo {author} {\bibfnamefont {S.}~\bibnamefont
  {Acharya}} \emph {et~al.} (\bibinfo {collaboration} {ALICE}),\ }\href
  {\doibase 10.1016/j.physletb.2019.134822} {\bibfield  {journal} {\bibinfo
  {journal} {Phys. Lett. B}\ }\textbf {\bibinfo {volume} {797}},\ \bibinfo
  {pages} {134822} (\bibinfo {year} {2019}{\natexlab{c}})},\ \Eprint
  {http://arxiv.org/abs/1905.07209} {arXiv:1905.07209 [nucl-ex]} \BibitemShut
  {NoStop}%
\bibitem [{\citenamefont {Acharya}\ \emph
  {et~al.}(2020{\natexlab{b}})\citenamefont {Acharya} \emph
  {et~al.}}]{ALICE:2019gcn}%
  \BibitemOpen
  \bibfield  {author} {\bibinfo {author} {\bibfnamefont {S.}~\bibnamefont
  {Acharya}} \emph {et~al.} (\bibinfo {collaboration} {ALICE}),\ }\href
  {\doibase 10.1103/PhysRevLett.124.092301} {\bibfield  {journal} {\bibinfo
  {journal} {Phys. Rev. Lett.}\ }\textbf {\bibinfo {volume} {124}},\ \bibinfo
  {pages} {092301} (\bibinfo {year} {2020}{\natexlab{b}})},\ \Eprint
  {http://arxiv.org/abs/1905.13470} {arXiv:1905.13470 [nucl-ex]} \BibitemShut
  {NoStop}%
\bibitem [{\citenamefont {Acharya}\ \emph
  {et~al.}(2019{\natexlab{d}})\citenamefont {Acharya} \emph
  {et~al.}}]{ALICE:2019hdt}%
  \BibitemOpen
  \bibfield  {author} {\bibinfo {author} {\bibfnamefont {S.}~\bibnamefont
  {Acharya}} \emph {et~al.} (\bibinfo {collaboration} {ALICE}),\ }\href
  {\doibase 10.1103/PhysRevLett.123.112002} {\bibfield  {journal} {\bibinfo
  {journal} {Phys. Rev. Lett.}\ }\textbf {\bibinfo {volume} {123}},\ \bibinfo
  {pages} {112002} (\bibinfo {year} {2019}{\natexlab{d}})},\ \Eprint
  {http://arxiv.org/abs/1904.12198} {arXiv:1904.12198 [nucl-ex]} \BibitemShut
  {NoStop}%
\bibitem [{\citenamefont {Acharya}\ \emph
  {et~al.}(2020{\natexlab{c}})\citenamefont {Acharya} \emph
  {et~al.}}]{ALICE:2019buq}%
  \BibitemOpen
  \bibfield  {author} {\bibinfo {author} {\bibfnamefont {S.}~\bibnamefont
  {Acharya}} \emph {et~al.} (\bibinfo {collaboration} {ALICE}),\ }\href
  {\doibase 10.1016/j.physletb.2020.135419} {\bibfield  {journal} {\bibinfo
  {journal} {Phys. Lett. B}\ }\textbf {\bibinfo {volume} {805}},\ \bibinfo
  {pages} {135419} (\bibinfo {year} {2020}{\natexlab{c}})},\ \Eprint
  {http://arxiv.org/abs/1910.14407} {arXiv:1910.14407 [nucl-ex]} \BibitemShut
  {NoStop}%
\bibitem [{\citenamefont {Acharya}\ \emph
  {et~al.}(2021{\natexlab{a}})\citenamefont {Acharya} \emph
  {et~al.}}]{ALICE:2020wvi}%
  \BibitemOpen
  \bibfield  {author} {\bibinfo {author} {\bibfnamefont {S.}~\bibnamefont
  {Acharya}} \emph {et~al.} (\bibinfo {collaboration} {ALICE}),\ }\href
  {\doibase 10.1103/PhysRevC.103.055201} {\bibfield  {journal} {\bibinfo
  {journal} {Phys. Rev. C}\ }\textbf {\bibinfo {volume} {103}},\ \bibinfo
  {pages} {055201} (\bibinfo {year} {2021}{\natexlab{a}})},\ \Eprint
  {http://arxiv.org/abs/2005.11124} {arXiv:2005.11124 [nucl-ex]} \BibitemShut
  {NoStop}%
\bibitem [{\citenamefont {{{ALICE Collaboration}}}(2020)}]{ALICE:2020mfd}%
  \BibitemOpen
  \bibfield  {author} {\bibinfo {author} {\bibnamefont {{{ALICE
  Collaboration}}}} (\bibinfo {collaboration} {ALICE}),\ }\href {\doibase
  10.1038/s41586-020-3001-6} {\bibfield  {journal} {\bibinfo  {journal}
  {Nature}\ }\textbf {\bibinfo {volume} {588}},\ \bibinfo {pages} {232}
  (\bibinfo {year} {2020})},\ \bibinfo {note} {[Erratum: Nature 590, E13
  (2021)]},\ \Eprint {http://arxiv.org/abs/2005.11495} {arXiv:2005.11495
  [nucl-ex]} \BibitemShut {NoStop}%
\bibitem [{\citenamefont {Acharya}\ \emph
  {et~al.}(2021{\natexlab{b}})\citenamefont {Acharya} \emph
  {et~al.}}]{ALICE:2020mkb}%
  \BibitemOpen
  \bibfield  {author} {\bibinfo {author} {\bibfnamefont {S.}~\bibnamefont
  {Acharya}} \emph {et~al.} (\bibinfo {collaboration} {ALICE}),\ }\href
  {\doibase 10.1016/j.physletb.2020.136030} {\bibfield  {journal} {\bibinfo
  {journal} {Phys. Lett. B}\ }\textbf {\bibinfo {volume} {813}},\ \bibinfo
  {pages} {136030} (\bibinfo {year} {2021}{\natexlab{b}})},\ \Eprint
  {http://arxiv.org/abs/2007.08315} {arXiv:2007.08315 [nucl-ex]} \BibitemShut
  {NoStop}%
\bibitem [{\citenamefont {Acharya}\ \emph
  {et~al.}(2022{\natexlab{a}})\citenamefont {Acharya} \emph
  {et~al.}}]{ALICE:2021njx}%
  \BibitemOpen
  \bibfield  {author} {\bibinfo {author} {\bibfnamefont {S.}~\bibnamefont
  {Acharya}} \emph {et~al.} (\bibinfo {collaboration} {ALICE}),\ }\href
  {\doibase 10.1016/j.physletb.2022.137272} {\bibfield  {journal} {\bibinfo
  {journal} {Phys. Lett. B}\ }\textbf {\bibinfo {volume} {833}},\ \bibinfo
  {pages} {137272} (\bibinfo {year} {2022}{\natexlab{a}})},\ \Eprint
  {http://arxiv.org/abs/2104.04427} {arXiv:2104.04427 [nucl-ex]} \BibitemShut
  {NoStop}%
\bibitem [{\citenamefont {Acharya}\ \emph
  {et~al.}(2022{\natexlab{b}})\citenamefont {Acharya} \emph
  {et~al.}}]{ALICE:2021ovd}%
  \BibitemOpen
  \bibfield  {author} {\bibinfo {author} {\bibfnamefont {S.}~\bibnamefont
  {Acharya}} \emph {et~al.} (\bibinfo {collaboration} {ALICE}),\ }\href
  {\doibase 10.1016/j.physletb.2022.137335} {\bibfield  {journal} {\bibinfo
  {journal} {Phys. Lett. B}\ }\textbf {\bibinfo {volume} {833}},\ \bibinfo
  {pages} {137335} (\bibinfo {year} {2022}{\natexlab{b}})},\ \Eprint
  {http://arxiv.org/abs/2111.06611} {arXiv:2111.06611 [nucl-ex]} \BibitemShut
  {NoStop}%
\bibitem [{\citenamefont {Acharya}\ \emph
  {et~al.}(2022{\natexlab{c}})\citenamefont {Acharya} \emph
  {et~al.}}]{ALICE:2021cyj}%
  \BibitemOpen
  \bibfield  {author} {\bibinfo {author} {\bibfnamefont {S.}~\bibnamefont
  {Acharya}} \emph {et~al.} (\bibinfo {collaboration} {ALICE}),\ }\href
  {\doibase 10.1016/j.physletb.2022.137060} {\bibfield  {journal} {\bibinfo
  {journal} {Phys. Lett. B}\ }\textbf {\bibinfo {volume} {829}},\ \bibinfo
  {pages} {137060} (\bibinfo {year} {2022}{\natexlab{c}})},\ \Eprint
  {http://arxiv.org/abs/2105.05190} {arXiv:2105.05190 [nucl-ex]} \BibitemShut
  {NoStop}%
\bibitem [{\citenamefont {Acharya}\ \emph
  {et~al.}(2021{\natexlab{c}})\citenamefont {Acharya} \emph
  {et~al.}}]{ALICE:2021cpv}%
  \BibitemOpen
  \bibfield  {author} {\bibinfo {author} {\bibfnamefont {S.}~\bibnamefont
  {Acharya}} \emph {et~al.} (\bibinfo {collaboration} {ALICE}),\ }\href
  {\doibase 10.1103/PhysRevLett.127.172301} {\bibfield  {journal} {\bibinfo
  {journal} {Phys. Rev. Lett.}\ }\textbf {\bibinfo {volume} {127}},\ \bibinfo
  {pages} {172301} (\bibinfo {year} {2021}{\natexlab{c}})},\ \Eprint
  {http://arxiv.org/abs/2105.05578} {arXiv:2105.05578 [nucl-ex]} \BibitemShut
  {NoStop}%
\bibitem [{\citenamefont {Acharya}\ \emph
  {et~al.}(2023{\natexlab{a}})\citenamefont {Acharya} \emph
  {et~al.}}]{ALICE:2022yyh}%
  \BibitemOpen
  \bibfield  {author} {\bibinfo {author} {\bibfnamefont {S.}~\bibnamefont
  {Acharya}} \emph {et~al.} (\bibinfo {collaboration} {ALICE}),\ }\href
  {\doibase 10.1140/epjc/s10052-023-11476-0} {\bibfield  {journal} {\bibinfo
  {journal} {Eur. Phys. J. C}\ }\textbf {\bibinfo {volume} {83}},\ \bibinfo
  {pages} {340} (\bibinfo {year} {2023}{\natexlab{a}})},\ \Eprint
  {http://arxiv.org/abs/2205.15176} {arXiv:2205.15176 [nucl-ex]} \BibitemShut
  {NoStop}%
\bibitem [{\citenamefont {Acharya}\ \emph
  {et~al.}(2023{\natexlab{b}})\citenamefont {Acharya} \emph
  {et~al.}}]{ALICE:2022uso}%
  \BibitemOpen
  \bibfield  {author} {\bibinfo {author} {\bibfnamefont {S.}~\bibnamefont
  {Acharya}} \emph {et~al.} (\bibinfo {collaboration} {ALICE}),\ }\href
  {\doibase 10.1016/j.physletb.2022.137223} {\bibfield  {journal} {\bibinfo
  {journal} {Phys. Lett. B}\ }\textbf {\bibinfo {volume} {844}},\ \bibinfo
  {pages} {137223} (\bibinfo {year} {2023}{\natexlab{b}})},\ \Eprint
  {http://arxiv.org/abs/2204.10258} {arXiv:2204.10258 [nucl-ex]} \BibitemShut
  {NoStop}%
\bibitem [{\citenamefont {Acharya}\ \emph
  {et~al.}(2023{\natexlab{c}})\citenamefont {Acharya} \emph
  {et~al.}}]{ALICE:2022mxo}%
  \BibitemOpen
  \bibfield  {author} {\bibinfo {author} {\bibfnamefont {S.}~\bibnamefont
  {Acharya}} \emph {et~al.} (\bibinfo {collaboration} {ALICE}),\ }\href
  {\doibase 10.1103/PhysRevC.107.054904} {\bibfield  {journal} {\bibinfo
  {journal} {Phys. Rev. C}\ }\textbf {\bibinfo {volume} {107}},\ \bibinfo
  {pages} {054904} (\bibinfo {year} {2023}{\natexlab{c}})},\ \Eprint
  {http://arxiv.org/abs/2211.15194} {arXiv:2211.15194 [nucl-ex]} \BibitemShut
  {NoStop}%
\bibitem [{\citenamefont {Acharya}\ \emph
  {et~al.}(2023{\natexlab{d}})\citenamefont {Acharya} \emph
  {et~al.}}]{ALICE:2023wjz}%
  \BibitemOpen
  \bibfield  {author} {\bibinfo {author} {\bibfnamefont {S.}~\bibnamefont
  {Acharya}} \emph {et~al.} (\bibinfo {collaboration} {ALICE}),\ }\href
  {\doibase 10.1016/j.physletb.2023.138145} {\bibfield  {journal} {\bibinfo
  {journal} {Phys. Lett. B}\ }\textbf {\bibinfo {volume} {845}},\ \bibinfo
  {pages} {138145} (\bibinfo {year} {2023}{\natexlab{d}})},\ \Eprint
  {http://arxiv.org/abs/2305.19093} {arXiv:2305.19093 [nucl-ex]} \BibitemShut
  {NoStop}%
\bibitem [{\citenamefont {Acharya}\ \emph {et~al.}(2024)\citenamefont {Acharya}
  \emph {et~al.}}]{ALICE:2024bhk}%
  \BibitemOpen
  \bibfield  {author} {\bibinfo {author} {\bibfnamefont {S.}~\bibnamefont
  {Acharya}} \emph {et~al.} (\bibinfo {collaboration} {ALICE}),\ }\href
  {\doibase 10.1103/PhysRevD.110.032004} {\bibfield  {journal} {\bibinfo
  {journal} {Phys. Rev. D}\ }\textbf {\bibinfo {volume} {110}},\ \bibinfo
  {pages} {032004} (\bibinfo {year} {2024})},\ \Eprint
  {http://arxiv.org/abs/2401.13541} {arXiv:2401.13541 [nucl-ex]} \BibitemShut
  {NoStop}%
\bibitem [{\citenamefont {Acharya}\ \emph
  {et~al.}(2022{\natexlab{d}})\citenamefont {Acharya} \emph
  {et~al.}}]{ALICE:2022enj}%
  \BibitemOpen
  \bibfield  {author} {\bibinfo {author} {\bibfnamefont {S.}~\bibnamefont
  {Acharya}} \emph {et~al.} (\bibinfo {collaboration} {ALICE}),\ }\href
  {\doibase 10.1103/PhysRevD.106.052010} {\bibfield  {journal} {\bibinfo
  {journal} {Phys. Rev. D}\ }\textbf {\bibinfo {volume} {106}},\ \bibinfo
  {pages} {052010} (\bibinfo {year} {2022}{\natexlab{d}})},\ \Eprint
  {http://arxiv.org/abs/2201.05352} {arXiv:2201.05352 [nucl-ex]} \BibitemShut
  {NoStop}%
\bibitem [{\citenamefont {Kamiya}\ \emph {et~al.}(2022)\citenamefont {Kamiya},
  \citenamefont {Hyodo},\ and\ \citenamefont {Ohnishi}}]{Kamiya:2022thy}%
  \BibitemOpen
  \bibfield  {author} {\bibinfo {author} {\bibfnamefont {Y.}~\bibnamefont
  {Kamiya}}, \bibinfo {author} {\bibfnamefont {T.}~\bibnamefont {Hyodo}}, \
  and\ \bibinfo {author} {\bibfnamefont {A.}~\bibnamefont {Ohnishi}},\ }\href
  {\doibase 10.1140/epja/s10050-022-00782-y} {\bibfield  {journal} {\bibinfo
  {journal} {Eur. Phys. J. A}\ }\textbf {\bibinfo {volume} {58}},\ \bibinfo
  {pages} {131} (\bibinfo {year} {2022})},\ \Eprint
  {http://arxiv.org/abs/2203.13814} {arXiv:2203.13814 [hep-ph]} \BibitemShut
  {NoStop}%
\bibitem [{\citenamefont {Abreu}\ and\ \citenamefont
  {Torres-Rincon}(2025)}]{Abreu:2025jqy}%
  \BibitemOpen
  \bibfield  {author} {\bibinfo {author} {\bibfnamefont {L.~M.}\ \bibnamefont
  {Abreu}}\ and\ \bibinfo {author} {\bibfnamefont {J.~M.}\ \bibnamefont
  {Torres-Rincon}},\ }\href {\doibase 10.1103/vffb-kv1t} {\bibfield  {journal}
  {\bibinfo  {journal} {Phys. Rev. D}\ }\textbf {\bibinfo {volume} {112}},\
  \bibinfo {pages} {016003} (\bibinfo {year} {2025})},\ \Eprint
  {http://arxiv.org/abs/2503.13668} {arXiv:2503.13668 [hep-ph]} \BibitemShut
  {NoStop}%
\bibitem [{\citenamefont {Vidana}\ \emph {et~al.}(2023)\citenamefont {Vidana},
  \citenamefont {Feijoo}, \citenamefont {Albaladejo}, \citenamefont {Nieves},\
  and\ \citenamefont {Oset}}]{Vidana:2023olz}%
  \BibitemOpen
  \bibfield  {author} {\bibinfo {author} {\bibfnamefont {I.}~\bibnamefont
  {Vidana}}, \bibinfo {author} {\bibfnamefont {A.}~\bibnamefont {Feijoo}},
  \bibinfo {author} {\bibfnamefont {M.}~\bibnamefont {Albaladejo}}, \bibinfo
  {author} {\bibfnamefont {J.}~\bibnamefont {Nieves}}, \ and\ \bibinfo {author}
  {\bibfnamefont {E.}~\bibnamefont {Oset}},\ }\href {\doibase
  10.1016/j.physletb.2023.138201} {\bibfield  {journal} {\bibinfo  {journal}
  {Phys. Lett. B}\ }\textbf {\bibinfo {volume} {846}},\ \bibinfo {pages}
  {138201} (\bibinfo {year} {2023})},\ \Eprint
  {http://arxiv.org/abs/2303.06079} {arXiv:2303.06079 [hep-ph]} \BibitemShut
  {NoStop}%
\bibitem [{\citenamefont {Albaladejo}\ \emph {et~al.}(2024)\citenamefont
  {Albaladejo}, \citenamefont {Feijoo}, \citenamefont {Nieves}, \citenamefont
  {Oset},\ and\ \citenamefont {Vida\~na}}]{Albaladejo:2024lam}%
  \BibitemOpen
  \bibfield  {author} {\bibinfo {author} {\bibfnamefont {M.}~\bibnamefont
  {Albaladejo}}, \bibinfo {author} {\bibfnamefont {A.}~\bibnamefont {Feijoo}},
  \bibinfo {author} {\bibfnamefont {J.}~\bibnamefont {Nieves}}, \bibinfo
  {author} {\bibfnamefont {E.}~\bibnamefont {Oset}}, \ and\ \bibinfo {author}
  {\bibfnamefont {I.}~\bibnamefont {Vida\~na}},\ }\href {\doibase
  10.1103/PhysRevD.110.114052} {\bibfield  {journal} {\bibinfo  {journal}
  {Phys. Rev. D}\ }\textbf {\bibinfo {volume} {110}},\ \bibinfo {pages}
  {114052} (\bibinfo {year} {2024})},\ \Eprint
  {http://arxiv.org/abs/2410.08880} {arXiv:2410.08880 [hep-ph]} \BibitemShut
  {NoStop}%
\bibitem [{\citenamefont {Liu}\ \emph {et~al.}(2024)\citenamefont {Liu},
  \citenamefont {Lu}, \citenamefont {Liu},\ and\ \citenamefont
  {Geng}}]{Liu:2024nac}%
  \BibitemOpen
  \bibfield  {author} {\bibinfo {author} {\bibfnamefont {Z.-W.}\ \bibnamefont
  {Liu}}, \bibinfo {author} {\bibfnamefont {J.-X.}\ \bibnamefont {Lu}},
  \bibinfo {author} {\bibfnamefont {M.-Z.}\ \bibnamefont {Liu}}, \ and\
  \bibinfo {author} {\bibfnamefont {L.-S.}\ \bibnamefont {Geng}},\ }\href@noop
  {} {\  (\bibinfo {year} {2024})},\ \Eprint {http://arxiv.org/abs/2404.18607}
  {arXiv:2404.18607 [hep-ph]} \BibitemShut {NoStop}%
\bibitem [{\citenamefont {Liu}\ \emph {et~al.}(2023{\natexlab{a}})\citenamefont
  {Liu}, \citenamefont {Lu}, \citenamefont {Liu},\ and\ \citenamefont
  {Geng}}]{Liu:2023wfo}%
  \BibitemOpen
  \bibfield  {author} {\bibinfo {author} {\bibfnamefont {Z.-W.}\ \bibnamefont
  {Liu}}, \bibinfo {author} {\bibfnamefont {J.-X.}\ \bibnamefont {Lu}},
  \bibinfo {author} {\bibfnamefont {M.-Z.}\ \bibnamefont {Liu}}, \ and\
  \bibinfo {author} {\bibfnamefont {L.-S.}\ \bibnamefont {Geng}},\ }\href
  {\doibase 10.1103/PhysRevD.108.L031503} {\bibfield  {journal} {\bibinfo
  {journal} {Phys. Rev. D}\ }\textbf {\bibinfo {volume} {108}},\ \bibinfo
  {pages} {L031503} (\bibinfo {year} {2023}{\natexlab{a}})},\ \Eprint
  {http://arxiv.org/abs/2305.19048} {arXiv:2305.19048 [hep-ph]} \BibitemShut
  {NoStop}%
\bibitem [{\citenamefont {Albaladejo}\ \emph {et~al.}(2023)\citenamefont
  {Albaladejo}, \citenamefont {Nieves},\ and\ \citenamefont
  {Ruiz-Arriola}}]{Albaladejo:2023pzq}%
  \BibitemOpen
  \bibfield  {author} {\bibinfo {author} {\bibfnamefont {M.}~\bibnamefont
  {Albaladejo}}, \bibinfo {author} {\bibfnamefont {J.}~\bibnamefont {Nieves}},
  \ and\ \bibinfo {author} {\bibfnamefont {E.}~\bibnamefont {Ruiz-Arriola}},\
  }\href {\doibase 10.1103/PhysRevD.108.014020} {\bibfield  {journal} {\bibinfo
   {journal} {Phys. Rev. D}\ }\textbf {\bibinfo {volume} {108}},\ \bibinfo
  {pages} {014020} (\bibinfo {year} {2023})},\ \Eprint
  {http://arxiv.org/abs/2304.03107} {arXiv:2304.03107 [hep-ph]} \BibitemShut
  {NoStop}%
\bibitem [{\citenamefont {Torres-Rincon}\ \emph {et~al.}(2023)\citenamefont
  {Torres-Rincon}, \citenamefont {Ramos},\ and\ \citenamefont
  {Tolos}}]{Torres-Rincon:2023qll}%
  \BibitemOpen
  \bibfield  {author} {\bibinfo {author} {\bibfnamefont {J.~M.}\ \bibnamefont
  {Torres-Rincon}}, \bibinfo {author} {\bibfnamefont {A.}~\bibnamefont
  {Ramos}}, \ and\ \bibinfo {author} {\bibfnamefont {L.}~\bibnamefont
  {Tolos}},\ }\href {\doibase 10.1103/PhysRevD.108.096008} {\bibfield
  {journal} {\bibinfo  {journal} {Phys. Rev. D}\ }\textbf {\bibinfo {volume}
  {108}},\ \bibinfo {pages} {096008} (\bibinfo {year} {2023})},\ \Eprint
  {http://arxiv.org/abs/2307.02102} {arXiv:2307.02102 [hep-ph]} \BibitemShut
  {NoStop}%
\bibitem [{\citenamefont {Liu}\ \emph {et~al.}(2023{\natexlab{b}})\citenamefont
  {Liu}, \citenamefont {Lu},\ and\ \citenamefont {Geng}}]{Liu:2023uly}%
  \BibitemOpen
  \bibfield  {author} {\bibinfo {author} {\bibfnamefont {Z.-W.}\ \bibnamefont
  {Liu}}, \bibinfo {author} {\bibfnamefont {J.-X.}\ \bibnamefont {Lu}}, \ and\
  \bibinfo {author} {\bibfnamefont {L.-S.}\ \bibnamefont {Geng}},\ }\href
  {\doibase 10.1103/PhysRevD.107.074019} {\bibfield  {journal} {\bibinfo
  {journal} {Phys. Rev. D}\ }\textbf {\bibinfo {volume} {107}},\ \bibinfo
  {pages} {074019} (\bibinfo {year} {2023}{\natexlab{b}})},\ \Eprint
  {http://arxiv.org/abs/2302.01046} {arXiv:2302.01046 [hep-ph]} \BibitemShut
  {NoStop}%
\bibitem [{\citenamefont {Ikeno}\ \emph {et~al.}(2023)\citenamefont {Ikeno},
  \citenamefont {Toledo},\ and\ \citenamefont {Oset}}]{Ikeno:2023ojl}%
  \BibitemOpen
  \bibfield  {author} {\bibinfo {author} {\bibfnamefont {N.}~\bibnamefont
  {Ikeno}}, \bibinfo {author} {\bibfnamefont {G.}~\bibnamefont {Toledo}}, \
  and\ \bibinfo {author} {\bibfnamefont {E.}~\bibnamefont {Oset}},\ }\href
  {\doibase 10.1016/j.physletb.2023.138281} {\bibfield  {journal} {\bibinfo
  {journal} {Phys. Lett. B}\ }\textbf {\bibinfo {volume} {847}},\ \bibinfo
  {pages} {138281} (\bibinfo {year} {2023})},\ \Eprint
  {http://arxiv.org/abs/2305.16431} {arXiv:2305.16431 [hep-ph]} \BibitemShut
  {NoStop}%
\bibitem [{\citenamefont {Khemchandani}\ \emph {et~al.}(2024)\citenamefont
  {Khemchandani}, \citenamefont {Abreu}, \citenamefont {Martinez~Torres},\ and\
  \citenamefont {Navarra}}]{Khemchandani:2023xup}%
  \BibitemOpen
  \bibfield  {author} {\bibinfo {author} {\bibfnamefont {K.~P.}\ \bibnamefont
  {Khemchandani}}, \bibinfo {author} {\bibfnamefont {L.~M.}\ \bibnamefont
  {Abreu}}, \bibinfo {author} {\bibfnamefont {A.}~\bibnamefont
  {Martinez~Torres}}, \ and\ \bibinfo {author} {\bibfnamefont {F.~S.}\
  \bibnamefont {Navarra}},\ }\href {\doibase 10.1103/PhysRevD.110.036008}
  {\bibfield  {journal} {\bibinfo  {journal} {Phys. Rev. D}\ }\textbf {\bibinfo
  {volume} {110}},\ \bibinfo {pages} {036008} (\bibinfo {year} {2024})},\
  \Eprint {http://arxiv.org/abs/2312.11811} {arXiv:2312.11811 [hep-ph]}
  \BibitemShut {NoStop}%
\bibitem [{\citenamefont {Liu}\ \emph {et~al.}(2025)\citenamefont {Liu},
  \citenamefont {Ge}, \citenamefont {Lu}, \citenamefont {Liu},\ and\
  \citenamefont {Geng}}]{Liu:2025oar}%
  \BibitemOpen
  \bibfield  {author} {\bibinfo {author} {\bibfnamefont {Z.-W.}\ \bibnamefont
  {Liu}}, \bibinfo {author} {\bibfnamefont {D.-L.}\ \bibnamefont {Ge}},
  \bibinfo {author} {\bibfnamefont {J.-X.}\ \bibnamefont {Lu}}, \bibinfo
  {author} {\bibfnamefont {M.-Z.}\ \bibnamefont {Liu}}, \ and\ \bibinfo
  {author} {\bibfnamefont {L.-S.}\ \bibnamefont {Geng}},\ }\href {\doibase
  10.1103/3bdh-blwh} {\bibfield  {journal} {\bibinfo  {journal} {Phys. Rev. D}\
  }\textbf {\bibinfo {volume} {112}},\ \bibinfo {pages} {054019} (\bibinfo
  {year} {2025})},\ \Eprint {http://arxiv.org/abs/2504.04853} {arXiv:2504.04853
  [hep-ph]} \BibitemShut {NoStop}%
\bibitem [{\citenamefont {Etminan}(2025)}]{Etminan:2025tiy}%
  \BibitemOpen
  \bibfield  {author} {\bibinfo {author} {\bibfnamefont {F.}~\bibnamefont
  {Etminan}},\ }\href@noop {} {\  (\bibinfo {year} {2025})},\ \Eprint
  {http://arxiv.org/abs/2506.14724} {arXiv:2506.14724 [nucl-th]} \BibitemShut
  {NoStop}%
\bibitem [{\citenamefont {Hofmann}\ and\ \citenamefont
  {Lutz}(2005)}]{Hofmann:2005sw}%
  \BibitemOpen
  \bibfield  {author} {\bibinfo {author} {\bibfnamefont {J.}~\bibnamefont
  {Hofmann}}\ and\ \bibinfo {author} {\bibfnamefont {M.~F.~M.}\ \bibnamefont
  {Lutz}},\ }\href {\doibase 10.1016/j.nuclphysa.2005.08.022} {\bibfield
  {journal} {\bibinfo  {journal} {Nucl. Phys. A}\ }\textbf {\bibinfo {volume}
  {763}},\ \bibinfo {pages} {90} (\bibinfo {year} {2005})},\ \Eprint
  {http://arxiv.org/abs/hep-ph/0507071} {arXiv:hep-ph/0507071} \BibitemShut
  {NoStop}%
\bibitem [{\citenamefont {Jimenez-Tejero}\ \emph {et~al.}(2009)\citenamefont
  {Jimenez-Tejero}, \citenamefont {Ramos},\ and\ \citenamefont
  {Vidana}}]{Jimenez-Tejero:2009cyn}%
  \BibitemOpen
  \bibfield  {author} {\bibinfo {author} {\bibfnamefont {C.~E.}\ \bibnamefont
  {Jimenez-Tejero}}, \bibinfo {author} {\bibfnamefont {A.}~\bibnamefont
  {Ramos}}, \ and\ \bibinfo {author} {\bibfnamefont {I.}~\bibnamefont
  {Vidana}},\ }\href {\doibase 10.1103/PhysRevC.80.055206} {\bibfield
  {journal} {\bibinfo  {journal} {Phys. Rev. C}\ }\textbf {\bibinfo {volume}
  {80}},\ \bibinfo {pages} {055206} (\bibinfo {year} {2009})},\ \Eprint
  {http://arxiv.org/abs/0907.5316} {arXiv:0907.5316 [hep-ph]} \BibitemShut
  {NoStop}%
\bibitem [{\citenamefont {Lutz}\ and\ \citenamefont
  {Korpa}(2006)}]{Lutz:2005vx}%
  \BibitemOpen
  \bibfield  {author} {\bibinfo {author} {\bibfnamefont {M.~F.~M.}\
  \bibnamefont {Lutz}}\ and\ \bibinfo {author} {\bibfnamefont {C.~L.}\
  \bibnamefont {Korpa}},\ }\href {\doibase 10.1016/j.physletb.2005.11.046}
  {\bibfield  {journal} {\bibinfo  {journal} {Phys. Lett. B}\ }\textbf
  {\bibinfo {volume} {633}},\ \bibinfo {pages} {43} (\bibinfo {year} {2006})},\
  \Eprint {http://arxiv.org/abs/nucl-th/0510006} {arXiv:nucl-th/0510006}
  \BibitemShut {NoStop}%
\bibitem [{\citenamefont {Mizutani}\ and\ \citenamefont
  {Ramos}(2006)}]{Mizutani:2006vq}%
  \BibitemOpen
  \bibfield  {author} {\bibinfo {author} {\bibfnamefont {T.}~\bibnamefont
  {Mizutani}}\ and\ \bibinfo {author} {\bibfnamefont {A.}~\bibnamefont
  {Ramos}},\ }\href {\doibase 10.1103/PhysRevC.74.065201} {\bibfield  {journal}
  {\bibinfo  {journal} {Phys. Rev. C}\ }\textbf {\bibinfo {volume} {74}},\
  \bibinfo {pages} {065201} (\bibinfo {year} {2006})},\ \Eprint
  {http://arxiv.org/abs/hep-ph/0607257} {arXiv:hep-ph/0607257} \BibitemShut
  {NoStop}%
\bibitem [{\citenamefont {Garcia-Recio}\ \emph {et~al.}(2009)\citenamefont
  {Garcia-Recio}, \citenamefont {Magas}, \citenamefont {Mizutani},
  \citenamefont {Nieves}, \citenamefont {Ramos}, \citenamefont {Salcedo},\ and\
  \citenamefont {Tolos}}]{Garcia-Recio:2008rjt}%
  \BibitemOpen
  \bibfield  {author} {\bibinfo {author} {\bibfnamefont {C.}~\bibnamefont
  {Garcia-Recio}}, \bibinfo {author} {\bibfnamefont {V.~K.}\ \bibnamefont
  {Magas}}, \bibinfo {author} {\bibfnamefont {T.}~\bibnamefont {Mizutani}},
  \bibinfo {author} {\bibfnamefont {J.}~\bibnamefont {Nieves}}, \bibinfo
  {author} {\bibfnamefont {A.}~\bibnamefont {Ramos}}, \bibinfo {author}
  {\bibfnamefont {L.~L.}\ \bibnamefont {Salcedo}}, \ and\ \bibinfo {author}
  {\bibfnamefont {L.}~\bibnamefont {Tolos}},\ }\href {\doibase
  10.1103/PhysRevD.79.054004} {\bibfield  {journal} {\bibinfo  {journal} {Phys.
  Rev. D}\ }\textbf {\bibinfo {volume} {79}},\ \bibinfo {pages} {054004}
  (\bibinfo {year} {2009})},\ \Eprint {http://arxiv.org/abs/0807.2969}
  {arXiv:0807.2969 [hep-ph]} \BibitemShut {NoStop}%
\bibitem [{\citenamefont {Haidenbauer}\ \emph {et~al.}(2011)\citenamefont
  {Haidenbauer}, \citenamefont {Krein}, \citenamefont {Meissner},\ and\
  \citenamefont {Tolos}}]{Haidenbauer:2010ch}%
  \BibitemOpen
  \bibfield  {author} {\bibinfo {author} {\bibfnamefont {J.}~\bibnamefont
  {Haidenbauer}}, \bibinfo {author} {\bibfnamefont {G.}~\bibnamefont {Krein}},
  \bibinfo {author} {\bibfnamefont {U.-G.}\ \bibnamefont {Meissner}}, \ and\
  \bibinfo {author} {\bibfnamefont {L.}~\bibnamefont {Tolos}},\ }\href
  {\doibase 10.1140/epja/i2011-11018-3} {\bibfield  {journal} {\bibinfo
  {journal} {Eur. Phys. J. A}\ }\textbf {\bibinfo {volume} {47}},\ \bibinfo
  {pages} {18} (\bibinfo {year} {2011})},\ \Eprint
  {http://arxiv.org/abs/1008.3794} {arXiv:1008.3794 [nucl-th]} \BibitemShut
  {NoStop}%
\bibitem [{\citenamefont {Haidenbauer}\ \emph {et~al.}(2007)\citenamefont
  {Haidenbauer}, \citenamefont {Krein}, \citenamefont {Meissner},\ and\
  \citenamefont {Sibirtsev}}]{Haidenbauer:2007jq}%
  \BibitemOpen
  \bibfield  {author} {\bibinfo {author} {\bibfnamefont {J.}~\bibnamefont
  {Haidenbauer}}, \bibinfo {author} {\bibfnamefont {G.}~\bibnamefont {Krein}},
  \bibinfo {author} {\bibfnamefont {U.-G.}\ \bibnamefont {Meissner}}, \ and\
  \bibinfo {author} {\bibfnamefont {A.}~\bibnamefont {Sibirtsev}},\ }\href
  {\doibase 10.1140/epja/i2007-10417-3} {\bibfield  {journal} {\bibinfo
  {journal} {Eur. Phys. J. A}\ }\textbf {\bibinfo {volume} {33}},\ \bibinfo
  {pages} {107} (\bibinfo {year} {2007})},\ \Eprint
  {http://arxiv.org/abs/0704.3668} {arXiv:0704.3668 [nucl-th]} \BibitemShut
  {NoStop}%
\bibitem [{\citenamefont {Fontoura}\ \emph {et~al.}(2013)\citenamefont
  {Fontoura}, \citenamefont {Krein},\ and\ \citenamefont
  {Vizcarra}}]{Fontoura:2012mz}%
  \BibitemOpen
  \bibfield  {author} {\bibinfo {author} {\bibfnamefont {C.~E.}\ \bibnamefont
  {Fontoura}}, \bibinfo {author} {\bibfnamefont {G.}~\bibnamefont {Krein}}, \
  and\ \bibinfo {author} {\bibfnamefont {V.~E.}\ \bibnamefont {Vizcarra}},\
  }\href {\doibase 10.1103/PhysRevC.87.025206} {\bibfield  {journal} {\bibinfo
  {journal} {Phys. Rev. C}\ }\textbf {\bibinfo {volume} {87}},\ \bibinfo
  {pages} {025206} (\bibinfo {year} {2013})},\ \Eprint
  {http://arxiv.org/abs/1208.4058} {arXiv:1208.4058 [nucl-th]} \BibitemShut
  {NoStop}%
\bibitem [{\citenamefont {Yamaguchi}\ \emph {et~al.}(2011)\citenamefont
  {Yamaguchi}, \citenamefont {Ohkoda}, \citenamefont {Yasui},\ and\
  \citenamefont {Hosaka}}]{Yamaguchi:2011xb}%
  \BibitemOpen
  \bibfield  {author} {\bibinfo {author} {\bibfnamefont {Y.}~\bibnamefont
  {Yamaguchi}}, \bibinfo {author} {\bibfnamefont {S.}~\bibnamefont {Ohkoda}},
  \bibinfo {author} {\bibfnamefont {S.}~\bibnamefont {Yasui}}, \ and\ \bibinfo
  {author} {\bibfnamefont {A.}~\bibnamefont {Hosaka}},\ }\href {\doibase
  10.1103/PhysRevD.84.014032} {\bibfield  {journal} {\bibinfo  {journal} {Phys.
  Rev. D}\ }\textbf {\bibinfo {volume} {84}},\ \bibinfo {pages} {014032}
  (\bibinfo {year} {2011})},\ \Eprint {http://arxiv.org/abs/1105.0734}
  {arXiv:1105.0734 [hep-ph]} \BibitemShut {NoStop}%
\bibitem [{\citenamefont {Oller}\ and\ \citenamefont
  {Meissner}(2001)}]{Oller:2000fj}%
  \BibitemOpen
  \bibfield  {author} {\bibinfo {author} {\bibfnamefont {J.~A.}\ \bibnamefont
  {Oller}}\ and\ \bibinfo {author} {\bibfnamefont {U.~G.}\ \bibnamefont
  {Meissner}},\ }\href {\doibase 10.1016/S0370-2693(01)00078-8} {\bibfield
  {journal} {\bibinfo  {journal} {Phys. Lett. B}\ }\textbf {\bibinfo {volume}
  {500}},\ \bibinfo {pages} {263} (\bibinfo {year} {2001})},\ \Eprint
  {http://arxiv.org/abs/hep-ph/0011146} {arXiv:hep-ph/0011146} \BibitemShut
  {NoStop}%
\bibitem [{\citenamefont {Kawarabayashi}\ and\ \citenamefont
  {Suzuki}(1966)}]{Kawarabayashi:1966kd}%
  \BibitemOpen
  \bibfield  {author} {\bibinfo {author} {\bibfnamefont {K.}~\bibnamefont
  {Kawarabayashi}}\ and\ \bibinfo {author} {\bibfnamefont {M.}~\bibnamefont
  {Suzuki}},\ }\href {\doibase 10.1103/PhysRevLett.16.255} {\bibfield
  {journal} {\bibinfo  {journal} {Phys. Rev. Lett.}\ }\textbf {\bibinfo
  {volume} {16}},\ \bibinfo {pages} {255} (\bibinfo {year} {1966})}\BibitemShut
  {NoStop}%
\bibitem [{\citenamefont {Riazuddin}\ and\ \citenamefont
  {Fayyazuddin}(1966)}]{Riazuddin:1966sw}%
  \BibitemOpen
  \bibfield  {author} {\bibinfo {author} {\bibnamefont {Riazuddin}}\ and\
  \bibinfo {author} {\bibnamefont {Fayyazuddin}},\ }\href {\doibase
  10.1103/PhysRev.147.1071} {\bibfield  {journal} {\bibinfo  {journal} {Phys.
  Rev.}\ }\textbf {\bibinfo {volume} {147}},\ \bibinfo {pages} {1071} (\bibinfo
  {year} {1966})}\BibitemShut {NoStop}%
\bibitem [{\citenamefont {Koonin}(1977)}]{KOONIN197743}%
  \BibitemOpen
  \bibfield  {author} {\bibinfo {author} {\bibfnamefont {S.~E.}\ \bibnamefont
  {Koonin}},\ }\href {\doibase https://doi.org/10.1016/0370-2693(77)90340-9}
  {\bibfield  {journal} {\bibinfo  {journal} {Physics Letters B}\ }\textbf
  {\bibinfo {volume} {70}},\ \bibinfo {pages} {43} (\bibinfo {year}
  {1977})}\BibitemShut {NoStop}%
\bibitem [{\citenamefont {Pratt}\ \emph {et~al.}(1990)\citenamefont {Pratt},
  \citenamefont {Cs\"org\ifmmode~\mbox{\H{o}}\else \H{o}\fi{}},\ and\
  \citenamefont {Zim\'anyi}}]{PhysRevC.42.2646}%
  \BibitemOpen
  \bibfield  {author} {\bibinfo {author} {\bibfnamefont {S.}~\bibnamefont
  {Pratt}}, \bibinfo {author} {\bibfnamefont {T.}~\bibnamefont
  {Cs\"org\ifmmode~\mbox{\H{o}}\else \H{o}\fi{}}}, \ and\ \bibinfo {author}
  {\bibfnamefont {J.}~\bibnamefont {Zim\'anyi}},\ }\href {\doibase
  10.1103/PhysRevC.42.2646} {\bibfield  {journal} {\bibinfo  {journal} {Phys.
  Rev. C}\ }\textbf {\bibinfo {volume} {42}},\ \bibinfo {pages} {2646}
  (\bibinfo {year} {1990})}\BibitemShut {NoStop}%
\bibitem [{\citenamefont {Gmitro}\ \emph {et~al.}(1986)\citenamefont {Gmitro},
  \citenamefont {Kvasil}, \citenamefont {Lednicky},\ and\ \citenamefont
  {Lyuboshits}}]{Gmitro:1986ay}%
  \BibitemOpen
  \bibfield  {author} {\bibinfo {author} {\bibfnamefont {M.}~\bibnamefont
  {Gmitro}}, \bibinfo {author} {\bibfnamefont {J.}~\bibnamefont {Kvasil}},
  \bibinfo {author} {\bibfnamefont {R.}~\bibnamefont {Lednicky}}, \ and\
  \bibinfo {author} {\bibfnamefont {V.~L.}\ \bibnamefont {Lyuboshits}},\ }\href
  {\doibase 10.1007/BF01598029} {\bibfield  {journal} {\bibinfo  {journal}
  {Czech. J. Phys. B}\ }\textbf {\bibinfo {volume} {36}},\ \bibinfo {pages}
  {1281} (\bibinfo {year} {1986})}\BibitemShut {NoStop}%
\bibitem [{\citenamefont {Torres-Rincon}\ \emph {et~al.}(2025)\citenamefont
  {Torres-Rincon}, \citenamefont {Ramos},\ and\ \citenamefont
  {Ruf{\'\i}}}]{Torres-Rincon:2024znb}%
  \BibitemOpen
  \bibfield  {author} {\bibinfo {author} {\bibfnamefont {J.~M.}\ \bibnamefont
  {Torres-Rincon}}, \bibinfo {author} {\bibfnamefont {A.}~\bibnamefont
  {Ramos}}, \ and\ \bibinfo {author} {\bibfnamefont {J.}~\bibnamefont
  {Ruf{\'\i}}},\ }\href {\doibase 10.1103/PhysRevC.111.044906} {\bibfield
  {journal} {\bibinfo  {journal} {Phys. Rev. C}\ }\textbf {\bibinfo {volume}
  {111}},\ \bibinfo {pages} {044906} (\bibinfo {year} {2025})},\ \Eprint
  {http://arxiv.org/abs/2410.23853} {arXiv:2410.23853 [nucl-th]} \BibitemShut
  {NoStop}%
\bibitem [{\citenamefont {Lednicky}\ and\ \citenamefont
  {Lyuboshits}(1982)}]{lednicky1982effect}%
  \BibitemOpen
  \bibfield  {author} {\bibinfo {author} {\bibfnamefont {R.}~\bibnamefont
  {Lednicky}}\ and\ \bibinfo {author} {\bibfnamefont {V.}~\bibnamefont
  {Lyuboshits}},\ }\href@noop {} {\bibfield  {journal} {\bibinfo  {journal}
  {Sov. J. Nucl. Phys.(Engl. Transl.);(United States)}\ }\textbf {\bibinfo
  {volume} {35}} (\bibinfo {year} {1982})}\BibitemShut {NoStop}%
\bibitem [{\citenamefont {Vovchenko}\ and\ \citenamefont
  {Stoecker}(2019)}]{Vovchenko:2019pjl}%
  \BibitemOpen
  \bibfield  {author} {\bibinfo {author} {\bibfnamefont {V.}~\bibnamefont
  {Vovchenko}}\ and\ \bibinfo {author} {\bibfnamefont {H.}~\bibnamefont
  {Stoecker}},\ }\href {\doibase 10.1016/j.cpc.2019.06.024} {\bibfield
  {journal} {\bibinfo  {journal} {Comput. Phys. Commun.}\ }\textbf {\bibinfo
  {volume} {244}},\ \bibinfo {pages} {295} (\bibinfo {year} {2019})},\ \Eprint
  {http://arxiv.org/abs/1901.05249} {arXiv:1901.05249 [nucl-th]} \BibitemShut
  {NoStop}%
\bibitem [{\citenamefont {Encarnaci{\'o}n}\ \emph {et~al.}(2025)\citenamefont
  {Encarnaci{\'o}n}, \citenamefont {Feijoo}, \citenamefont {Sarti},\ and\
  \citenamefont {Ramos}}]{Encarnacion:2024jge}%
  \BibitemOpen
  \bibfield  {author} {\bibinfo {author} {\bibfnamefont {P.}~\bibnamefont
  {Encarnaci{\'o}n}}, \bibinfo {author} {\bibfnamefont {A.}~\bibnamefont
  {Feijoo}}, \bibinfo {author} {\bibfnamefont {V.~M.}\ \bibnamefont {Sarti}}, \
  and\ \bibinfo {author} {\bibfnamefont {A.}~\bibnamefont {Ramos}},\ }\href
  {\doibase 10.1103/3ycr-vzmd} {\bibfield  {journal} {\bibinfo  {journal}
  {Phys. Rev. D}\ }\textbf {\bibinfo {volume} {111}},\ \bibinfo {pages}
  {114013} (\bibinfo {year} {2025})},\ \Eprint
  {http://arxiv.org/abs/2412.20880} {arXiv:2412.20880 [hep-ph]} \BibitemShut
  {NoStop}%
\bibitem [{\citenamefont {Letessier}\ and\ \citenamefont
  {Rafelski}(2022)}]{Letessier:2022fax}%
  \BibitemOpen
  \bibfield  {author} {\bibinfo {author} {\bibfnamefont {J.}~\bibnamefont
  {Letessier}}\ and\ \bibinfo {author} {\bibfnamefont {J.}~\bibnamefont
  {Rafelski}},\ }\href {\doibase 10.1017/9781009290753} {\emph {\bibinfo
  {title} {{Hadrons and Quark{\textendash}Gluon Plasma}}}}\ (\bibinfo
  {publisher} {Cambridge University Press},\ \bibinfo {year}
  {2022})\BibitemShut {NoStop}%
\bibitem [{\citenamefont {Sakai}\ \emph {et~al.}(2020)\citenamefont {Sakai},
  \citenamefont {Guo},\ and\ \citenamefont {Kubis}}]{SAKAI2020135623}%
  \BibitemOpen
  \bibfield  {author} {\bibinfo {author} {\bibfnamefont {S.}~\bibnamefont
  {Sakai}}, \bibinfo {author} {\bibfnamefont {F.-K.}\ \bibnamefont {Guo}}, \
  and\ \bibinfo {author} {\bibfnamefont {B.}~\bibnamefont {Kubis}},\ }\href
  {\doibase https://doi.org/10.1016/j.physletb.2020.135623} {\bibfield
  {journal} {\bibinfo  {journal} {Physics Letters B}\ }\textbf {\bibinfo
  {volume} {808}},\ \bibinfo {pages} {135623} (\bibinfo {year}
  {2020})}\BibitemShut {NoStop}%
\bibitem [{\citenamefont {Navas}\ \emph {et~al.}(2024)\citenamefont {Navas}
  \emph {et~al.}}]{ParticleDataGroup:2024cfk}%
  \BibitemOpen
  \bibfield  {author} {\bibinfo {author} {\bibfnamefont {S.}~\bibnamefont
  {Navas}} \emph {et~al.} (\bibinfo {collaboration} {Particle Data Group}),\
  }\href {\doibase 10.1103/PhysRevD.110.030001} {\bibfield  {journal} {\bibinfo
   {journal} {Phys. Rev. D}\ }\textbf {\bibinfo {volume} {110}},\ \bibinfo
  {pages} {030001} (\bibinfo {year} {2024})}\BibitemShut {NoStop}%
\bibitem [{\citenamefont {Jimenez-Tejero}\ \emph {et~al.}(2011)\citenamefont
  {Jimenez-Tejero}, \citenamefont {Ramos}, \citenamefont {Tolos},\ and\
  \citenamefont {Vidana}}]{Jimenez-Tejero:2011dif}%
  \BibitemOpen
  \bibfield  {author} {\bibinfo {author} {\bibfnamefont {C.~E.}\ \bibnamefont
  {Jimenez-Tejero}}, \bibinfo {author} {\bibfnamefont {A.}~\bibnamefont
  {Ramos}}, \bibinfo {author} {\bibfnamefont {L.}~\bibnamefont {Tolos}}, \ and\
  \bibinfo {author} {\bibfnamefont {I.}~\bibnamefont {Vidana}},\ }\href
  {\doibase 10.1103/PhysRevC.84.015208} {\bibfield  {journal} {\bibinfo
  {journal} {Phys. Rev. C}\ }\textbf {\bibinfo {volume} {84}},\ \bibinfo
  {pages} {015208} (\bibinfo {year} {2011})},\ \Eprint
  {http://arxiv.org/abs/1102.4786} {arXiv:1102.4786 [hep-ph]} \BibitemShut
  {NoStop}%
\bibitem [{\citenamefont {Epelbaum}\ \emph {et~al.}(2025)\citenamefont
  {Epelbaum}, \citenamefont {Heihoff}, \citenamefont {Mei{\ss}ner},\ and\
  \citenamefont {Tscherwon}}]{Epelbaum:2025aan}%
  \BibitemOpen
  \bibfield  {author} {\bibinfo {author} {\bibfnamefont {E.}~\bibnamefont
  {Epelbaum}}, \bibinfo {author} {\bibfnamefont {S.}~\bibnamefont {Heihoff}},
  \bibinfo {author} {\bibfnamefont {U.-G.}\ \bibnamefont {Mei{\ss}ner}}, \ and\
  \bibinfo {author} {\bibfnamefont {A.}~\bibnamefont {Tscherwon}},\ }\href@noop
  {} {\  (\bibinfo {year} {2025})},\ \Eprint {http://arxiv.org/abs/2504.08631}
  {arXiv:2504.08631 [nucl-th]} \BibitemShut {NoStop}%
\bibitem [{\citenamefont {Molina}\ and\ \citenamefont
  {Oset}(2025)}]{Molina:2025lzw}%
  \BibitemOpen
  \bibfield  {author} {\bibinfo {author} {\bibfnamefont {R.}~\bibnamefont
  {Molina}}\ and\ \bibinfo {author} {\bibfnamefont {E.}~\bibnamefont {Oset}},\
  }\href {\doibase 10.1103/rst4-rkmm} {\bibfield  {journal} {\bibinfo
  {journal} {Phys. Rev. D}\ }\textbf {\bibinfo {volume} {112}},\ \bibinfo
  {pages} {096006} (\bibinfo {year} {2025})},\ \Eprint
  {http://arxiv.org/abs/2506.03669} {arXiv:2506.03669 [hep-ph]} \BibitemShut
  {NoStop}%
\bibitem [{\citenamefont {Roy~Chowdhury}(2025)}]{RoyChowdhury:2025xhv}%
  \BibitemOpen
  \bibfield  {author} {\bibinfo {author} {\bibfnamefont {P.}~\bibnamefont
  {Roy~Chowdhury}},\ }\href {\doibase 10.1051/epjconf/202531604012} {\bibfield
  {journal} {\bibinfo  {journal} {EPJ Web Conf.}\ }\textbf {\bibinfo {volume}
  {316}},\ \bibinfo {pages} {04012} (\bibinfo {year} {2025})}\BibitemShut
  {NoStop}%
\bibitem [{\citenamefont {Joachain}(1975)}]{joachain1975quantum}%
  \BibitemOpen
  \bibfield  {author} {\bibinfo {author} {\bibfnamefont {C.}~\bibnamefont
  {Joachain}},\ }\href {https://books.google.es/books?id=Zs3vAAAAMAAJ} {\emph
  {\bibinfo {title} {Quantum Collision Theory}}}\ (\bibinfo  {publisher}
  {North-Holland Publishing Company},\ \bibinfo {year} {1975})\BibitemShut
  {NoStop}%
\bibitem [{\citenamefont {Holzenkamp}\ \emph {et~al.}(1989)\citenamefont
  {Holzenkamp}, \citenamefont {Holinde},\ and\ \citenamefont
  {Speth}}]{Holzenkamp:1989tq}%
  \BibitemOpen
  \bibfield  {author} {\bibinfo {author} {\bibfnamefont {B.}~\bibnamefont
  {Holzenkamp}}, \bibinfo {author} {\bibfnamefont {K.}~\bibnamefont {Holinde}},
  \ and\ \bibinfo {author} {\bibfnamefont {J.}~\bibnamefont {Speth}},\ }\href
  {\doibase 10.1016/0375-9474(89)90223-6} {\bibfield  {journal} {\bibinfo
  {journal} {Nucl. Phys. A}\ }\textbf {\bibinfo {volume} {500}},\ \bibinfo
  {pages} {485} (\bibinfo {year} {1989})}\BibitemShut {NoStop}%
\end{thebibliography}%
\bibliographystyle{apsrev4-1}

\end{document}